%% file: pasa-main.tex
\documentclass[
  journal=pasa,
  manuscript=research-paper, 
  year=2022,
  volume=,
]{cup-journal}

\usepackage{microtype,siunitx,booktabs}
\usepackage{amsmath}    
\usepackage{amssymb}
\usepackage{graphicx, xcolor,xspace}   
\usepackage{soul}
\usepackage[breaklinks=true,colorlinks=true,allcolors=blue]{hyperref}
\usepackage{stfloats}
\usepackage{hhline}
\usepackage{placeins}
\usepackage{threeparttable}
\title{DeepGlow: an efficient neural-network emulator of physical afterglow models for gamma-ray bursts and gravitational-wave events}

\author{Oliver~M.~Boersma}
\affiliation{
Anton Pannekoek Institute, University of Amsterdam, Postbus 94249, 1090 GE Amsterdam, The Netherlands}
\alsoaffiliation{
ASTRON, the Netherlands Institute for Radio Astronomy, Oude Hoogeveensedijk 4,7991 PD Dwingeloo, The Netherlands}
\email[O. M. Boersma]{o.m.boersma@uva.nl}
\author{Joeri~van~Leeuwen}
\affiliation{
ASTRON, the Netherlands Institute for Radio Astronomy, Oude Hoogeveensedijk 4,7991 PD Dwingeloo, The Netherlands}

\doi{}

\received {dd Mmm YYYY}
\revised  {dd Mmm YYYY}
\accepted {dd Mmm YYYY}
\published{dd Mmm YYYY}


\begin{document}
\newcommand{\Aks}{\citetalias{aksulu_exploring_2022}}
\def\jvl#1{\textcolor{red}{\bf[#1 -- JVL]}\xspace}

\input{content}


\bibliography{ref_thesis}

\appendix

\input{appendices}

\end{document}

%% file: content.tex
\begin{abstract}
Gamma-ray bursts (GRBs) and double neutron-star merger gravitational wave events 
are followed by afterglows that shine from X-rays to radio, 
and these broadband transients are generally interpreted using analytical models.
Such models are relatively fast to execute, 
and thus easily allow estimates of the energy and geometry parameters of the blast wave, 
through many trial-and-error model calculations. 
One problem, however, is that such analytical models do not capture the underlying physical processes as well as
more realistic relativistic numerical hydrodynamic (RHD) simulations do. 
Ideally, those simulations are used for parameter estimation instead, but their computational cost makes this intractable. To this end, we present DeepGlow, a highly efficient neural network architecture trained to emulate a computationally costly RHD-based model of GRB afterglows, to within a few percent accuracy. As a first scientific application, we compare both the emulator and a different analytical model calibrated to RHD simulations, to estimate the parameters of a broadband GRB afterglow. We find consistent results between these two models, and also give further evidence for a stellar wind progenitor environment around this GRB source. DeepGlow fuses simulations that are otherwise too complex to execute over all parameters, to real broadband data of current and future GRB afterglows.
\end{abstract}

\section{INTRODUCTION}
\label{sec:int}
Bright, distant sources such as Active Galactic Nuclei and 
Gamma-ray bursts (GRBs) that are variable or transient, are powered by 
relativistic blast waves \citep{blandford_fluid_1976}.
Following the detection of the first GRB afterglow ~\citep[GRB970228;][]{costa_discovery_1997},
modeling of these expanding explosions has been given great attention in the literature.
The recent detection of the 
 short gamma-ray burst GRB170817A~\citep[e.g.,][]{abbott_gravitational_2017}
coincident 
with the gravitational-wave detection of binary neutron star merger
GW170817~\citep{ligo_scientific_collaboration_and_virgo_collaboration_gw170817_2017},
and the subsequent 
detection of a multi-frequency afterglow
\citep[e.g.][]{chornock_electromagnetic_2017,coulter_swope_2017,alexander_electromagnetic_2017,haggard_deep_2017,hallinan_radio_2017} have further heightened this interest in detecting afterglows \citep[e.g.][]{2021A&A...650A.131B}.

As the afterglows are now known to produce radio, optical and X-ray emission,  
various (semi-)analytical models have been developed to analyze this broadband 
data~\citep{wijers_shocked_1997,sari_spectra_1998,granot_shape_2002,2012MNRAS.427.1329L,Ryan_2020}. Such analytical models are relatively fast to execute and are thus easily applicable in parameter estimation studies where the model needs to be calculated many times over (e.g.,~\citealt{panaitescu_properties_2002}). These models do not fully capture the physical processes, however, encompassed in more realistic relativistic hydrodynamical (RHD) simulation approaches like \texttt{BOXFIT}~\citep{van_eerten_gamma-ray_2012}. \texttt{BOXFIT} is built on top of a series of 2D RHD jet simulations which describe the dynamics of the afterglow. \texttt{BOXFIT} then interpolates between the output of these simulations, saved in a large number of compressed snapshots at fixed times, and applies a linear radiative transfer approach to calculate spectra and light curves. This method works partly because of the scale invariance of jets with different energies or densities, as demonstrated in~\citet{van_eerten_gamma-ray_2012}. This makes it possible to compute the afterglow for arbitrary energy and densities from the saved simulation snapshots with specific energy and density. To calculate the afterglow flux, it is assumed that the dominant radiation is synchrotron radiation. The jet fluid, computed in the RHD simulations, is divided into small cells for which the broadband synchrotron emission is calculated. A large number of rays are passed through these fluid cells and the observed flux is obtained by integrating the emission over these rays using the linear radiative transfer equation~\citep{van_eerten_gamma-ray_2010} for each ray. While \texttt{BOXFIT} has been used to characterize GRB afterglow data (e.g.,~\citealt{higgins_detailed_2019}), its computational cost makes it an unattractive, resource- and energy-heavy approach for studies that, e.g., 
fit a large population of GRB afterglow data using sophisticated methods~\citep{aksulu_exploring_2022},
or simulate large numbers of afterglows to forecast how to infer binary and fireball parameters from future gravitational-wave  and electromagnetic detections \citep{2022A&A...664A.160B}.

In recent years, the use of machine learning techniques has exploded across the (astronomical) sciences to speed up various processes with high computational complexity (e.g.,~\citealt{schmit_emulation_2018,kasim_building_2021,kerzendorf_dalek_2021}). Specifically, deep learning and neural network (NN) methods have been used extensively owing to their ability to accurately replicate highly non-linear relations between input and output data~(e.g.,~\citealt{hornik_multilayer_1989,cybenko_approximation_1989}). Furthermore, after training the NNs, they are usually quick to execute because only a relative small number of computational steps need to be performed, if they are not too large. This is further aided by the existence of optimized deep learning libraries like \texttt{TensorFlow}\footnote{\url{https://www.tensorflow.org/}}.

In this work, we use NNs to emulate the output of \texttt{BOXFIT} with small evaluation cost compared to running \texttt{BOXFIT} itself. We emulate both interstellar medium (ISM) and stellar wind like progenitor environments. We verify the accuracy of our NNs by comparing the output to real \texttt{BOXFIT} data and then test their validity by inferring the properties of the afterglow of GRB970508 (e.g.,~\citealt{panaitescu_properties_2002,yost_study_2003}) which has a large broadband dataset. We compare the results to a recent analytical model calibrated to \texttt{BOXFIT} simulation data (\citealt{ryan_gamma-ray_2015}, Ryan et al. in prep). The trained NNs are freely available in the \texttt{DeepGlow} Python package\footnote{\url{https://github.com/OMBoersma/DeepGlow}} and all code associated with the methodology in this work is present as well.

In Section~\ref{sec:meth}, we describe the methods used to generate the \texttt{BOXFIT} output training data and how we trained the NNs. We demonstrate the accuracy of \texttt{DeepGlow} in Section~\ref{sec:DGresults} and fit the broadband dataset of GRB970508 as a test case in Section~\ref{sec:testcase}. We conclude and look towards the future in Sections~\ref{sec:discussion} and~\ref{sec:conc}. 

\section{METHODS}
\label{sec:meth}
\subsection{Training data}
\label{subsec:trdata}

\begin{table}[h]
    \caption{GRB afterglow parameter distributions used to generate the training data. }
    \begin{tabular}{|ll|l|l|}
    \toprule
       \multicolumn{2}{|l|}{Parameter}  & Distribution & Range 
       \\
       \hline
       $\theta_0$         &(jet half angle) & Log-Uniform & $(0.01,0.5\pi)$ rad   \\
       $E_\mathrm{K,iso}$ &(explosion energy) & Log-Uniform & $(10^{50},10^{56})$ ergs  \\
       $n_{\mathrm{ref}}$ &(density)  & Log-Uniform & $(10^{-5},10^{3})$ $\mathrm{cm}^{-3}$  \\
       $\theta_\mathrm{obs}$ &(observer angle) &  Uniform & $(0.01,2 \times \theta_0)$ rad\\
       $p$ &(power-law index) & Uniform & $(2,3)$ \\
       $\epsilon_B$  &(magnetic energy fraction) &  Log-Uniform & $(10^{-10},1)$ \\
       $\epsilon_e$  &(electron energy fraction) &  Log-Uniform & $(10^{-10},1)$ \\
       $\nu_\mathrm{obs}$  &(observer frequency) &  Log-Uniform & $(10^{8},10^{19})$ Hz \\
       $t_\mathrm{obs}$  &(observer time) &  Fixed & $0.1$ d -  $1000$ d \\
       $\xi_N$  &(acc. electrons fraction) &  Fixed & $1$ \\
       $d_L$  &(luminosity distance) &  Fixed & $50$ Mpc \\
       $z$  &(redshift) &  Fixed & $0$ \\
    \bottomrule
    \end{tabular}
    \label{tab:paramdists}
\end{table}

Several sets of simulation data are available to be used with \texttt{BOXFIT}\footnote{\url{https://cosmo.nyu.edu/afterglowlibrary/boxfit2011.html}}. These differ in the progenitor environment and the Lorentz frame used to do the simulations. We used two sets of simulation data in this work, the "lab frame ISM environment" set and the "medium boost wind environment" set, and generated two corresponding sets of training data consisting of 200 000 light curves each. We employed the computing cluster of the Apertif Radio Transient System~\citep{van_leeuwen_apertif_2022} using 40 nodes each having 40 CPU cores. Generating the datasets took around a hundred thousand combined core hours.

To generate light curves with \texttt{BOXFIT}, ten GRB afterglow parameters must be specified: (i) the jet half opening angle $\theta_0$ in radians; (ii) the isotropic-equivalent explosion energy $E_\mathrm{K,iso}$ in ergs; (iii) the circumburst number density $n_{\mathrm{ref}}$ in $\mathrm{cm}^{-3}$; (iv) the observer angle $\theta_\mathrm{obs}$ in radians; (v) the index of the synchrotron power law slope of the accelerated electrons $p$; (vi) the fraction of thermal energy in the magnetic fields $\epsilon_B$; (vii) the fraction of thermal energy in the accelerated electrons $\epsilon_e$; (viii) the fraction of accelerated electrons $\xi_N$; (ix) the observer luminosity distance $d_L$ in cm; and (x) the cosmic redshift $z$. As the GRB afterglow flux scales straightforwardly with $\xi_N$, $d_L$, and $z$, we fixed those parameters to, in principle arbitrary, values of $1$, $50$ Mpc, and 0, respectively, when generating the training data. We chose the remaining parameters from broad log-uniform distributions except for $p$ which was uniformly distributed and $\theta_\mathrm{obs}$ which was distributed uniformly as well with a maximum $\theta_\mathrm{obs} < 2 \times \theta_0$. The parameter distributions are summarized in Table~\ref{tab:paramdists}.

We generated the light curves on a fixed observer time grid between $t_{\mathrm{obs},0} = 0.1$ days and $t_{\mathrm{obs,1}} = 1000$ days with 117 observer data points\footnote{The odd number is a consequence of the 39 threads (one thread is used as a controller thread) on each computing node available for parallelization.}. The amount of observer data points chosen is a trade-off between the time it takes to generate the training data and the resolution of the resulting light curves. 

\begin{figure}[t]
    \centering
    \includegraphics[width=\textwidth]{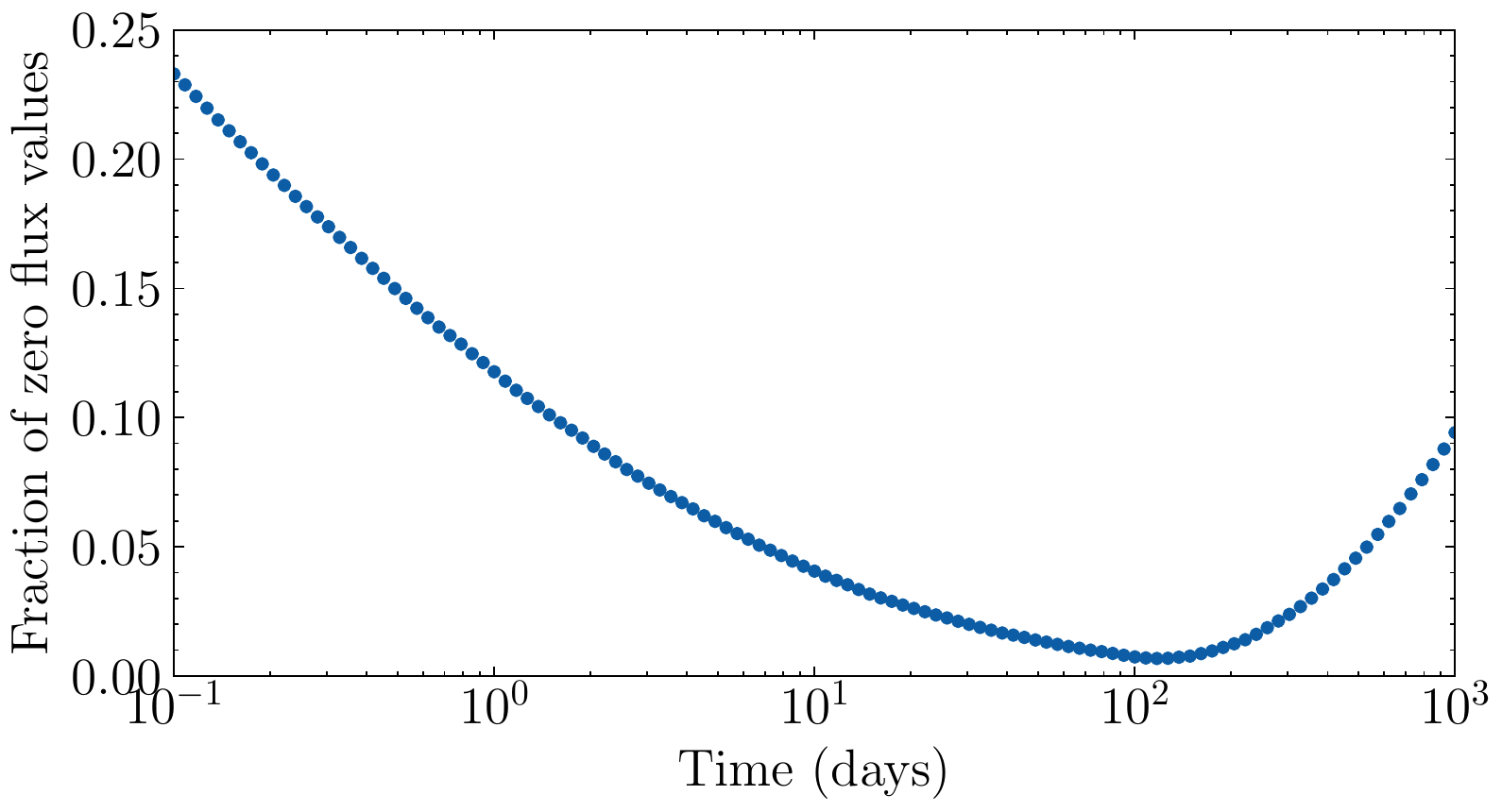}
    \caption{The fraction of light curves where the flux at a certain observer time is set to zero by \texttt{BOXFIT} when generating the training data. The dots correspond to the 117 datapoints generated for each light curve.}
    \label{fig:percentage_zeroflux}
\end{figure}

For 32\% (ISM environment) to 44\% (wind environment) of the generated light curves, the \texttt{BOXFIT} calculations did not cover the entire observer time range which resulted in zero flux values at observer times with no coverage. In Fig.~\ref{fig:percentage_zeroflux}, we show the fraction of zero flux values as a function of observer time. The flux at a certain observer time corresponds to the combined emission from a range of emission times~\citep{van_eerten_gamma-ray_2012}. If these times are not captured in the RHD simulations incorporated into \texttt{BOXFIT}, \texttt{BOXFIT} uses the Blandford–McKee (BM) solution~\citep{blandford_fluid_1976} at early times as a starting point, beginning with a fluid Lorentz factor of 300. 
Still, very early observer epochs are sometimes not covered by this BM solution either, because the thin shell of the shock at that time is not numerically resolved, or because the emission times occur before our asymptotic BM limit of Lorentz factor 300. For such observer times either the flux computed by \texttt{BOXFIT} is zero, or it rapidly drops off as only some of the emission times which are summed up can be calculated. For afterglow measurements of GRBs at such early times, \texttt{BOXFIT}, and by extension \texttt{DeepGlow}, is not a suitable model and should not be used. Furthermore, in the first few hours of GRB afterglow Swift~\citep{gehrels_swift_2004} data often a plateau phase is observed~(e.g.~\citealt{nousek_evidence_2006}), possibly due to a coasting or extended energy injection phase. These are not modeled in \texttt{BOXFIT}. Still, these afterglows will eventually evolve to a regime where \texttt{BOXFIT} is valid. It is also possible that very late observer times cannot be calculated by the \texttt{BOXFIT} simulations or the BM solution. In general, the broadband GRB afterglow datasets we are interested in modeling have measurements in the \texttt{BOXFIT} validity regime and the incomplete coverage is thus not an issue. Still, this does have an effect on the performance of \texttt{DeepGlow} as we will show in Section~\ref{sec:DGresults}. Overall, care must be taken when training the NNs on light curves with zero flux values and we return to this point in Section~\ref{subsec:NNarch}. 

We did not use a fixed grid of observer frequencies but chose a random $\nu_\mathrm{obs}$ for each light curve from a log-uniform distribution similar to the GRB afterglow parameters mentioned above. We picked a broad range in observer frequencies from $\nu_0 \ = \ 10^8$ Hz to $\nu_1 \ = \ 10^{19}$ Hz such that the trained NNs can be used to fit most broadband GRB afterglow datasets. 

Various resolution and radiation related parameters must be specified in \texttt{BOXFIT} which we left mostly to their default settings. We did adjust three parameters pertaining to the numerical resolution of the radiative transfer calculations in \texttt{BOXFIT}. More specifically, these three parameters correspond to (i,ii) the number of rays used in the radiative transfer steps in the radial and tangential direction; and (iii) the total number of simulation snapshots used. The first two parameters determine the number of rays for which the flux is calculated, while the third parameter sets the resolution along the rays\footnote{See the \texttt{BOXFIT} manual for more details.}. We set these to $500,\ 30,$ and $500$, respectively, which are lower values than the default settings, and this is again a trade-off between computing time and the quality of the light curve. From our testing, the numerical noise of the light curves with the mentioned values is similar to that of the default settings while saving about half of the computation time. 
\subsection{Neural Network setup and training}
\label{subsec:NNarch}
Each sample in our training data consists of eight input values ($\theta_0$, $E$, $n$, $p$, $\epsilon_B$, $\epsilon_e$, $\theta_\mathrm{obs}$ and $\nu_\mathrm{obs}$) and 117 output values corresponding to the observer data points. As a preprocessing step, we took the $\log_{10}$ of the input and output values (except for $p$ which has a limited range) and removed the mean and variance. GRB afterglow light curves generally follow a power-law decay~(e.g.,~\citealt{sari_spectra_1998}), thus the flux values at early observer times can be orders of magnitude larger than at late observer times. Taking the $\log_{10}$ of the flux values is a necessary step for the early time bins not to dominate the objective (loss) function for NN training. In log-space we used the \texttt{StandardScaler} in the \texttt{scikit-learn} Python package~\citep{scikit-learn} to standardize the individual time bins by subtracting the mean of the flux values and scaling to unit variance. The same steps were done for the input parameters. We also experimented with the \texttt{MinMaxScaler} but found this to produce slightly worse results. 

We modeled the relation between the resulting input and output values using a feed-forward NN (see e.g.,~\citealt{schmidhuber_deep_2015}, for an overview) in Keras/Tensorflow 2.9.1. One NN was trained for each of the two progenitor environments. The best architecture and hyperparameters of the NNs were obtained through a trial-and-error approach, i.e. manually searching the hyperparameter space. While, in the limit of finite training data, the reproduction will not be perfect, we aimed for \texttt{DeepGlow} to reproduce \texttt{BOXFIT} with an error which is generally well below the typical fractional GRB afterglow flux measurement error ($\approx 10  - 30 \%$ in the GRB970508 dataset). We experimented with the number of layers, the size of the layers, learning rate, activation function and batch size. Three large hidden layers consisting of 1000 neurons with a softplus activation function produced the best results. The output layer, consisting of 117 neurons, uses a linear activation function. We used Nesterov-adam~\citep{sutskever_importance_2013} as the optimizer with a cyclic learning rate (\textit{triangular2} policy,~\citealt{smith_cyclical_2015}) between $10^{-4}$ and $10^{-2}$. 
We employed 90\% of the training data to actually train the network, the remaining 10\% was used to the test the accuracy of the NNs afterwards. The data was fed through the network with a batch size of 128 for 2000 epochs and we selected the realization with the highest accuracy, i.e. lowest median fractional error compared to \texttt{BOXFIT} over the test dataset, from all epochs. In general, three to five equally-sized hidden layers with at least 200 neurons brought the reproduction error into our goal range. A cyclic learning rate schedule was also crucial in bringing the error down. Changing the activation function or the batch size had limited influence, however. We did not experiment with different optimizers.

NNs run the risk of overfitting on the training data and not generalizing well to unseen test data~(e.g.,~\citealt{caruana_overfitting_2000,ying_overview_2019}). Because of the uniform way in which the light curve data was generated through our Monte Carlo approach and the relatively large size of the training and test data, the risk of overfitting is low in our case. For reference, we compare the performance of the trained NNs on both the test dataset and training dataset in~\ref{app:extraNNacc}, where we show these are almost the same.

We chose the mean absolute error as our loss function;  using the mean squared error loss resulted in reproduction errors an order of magnitude worse. As mentioned in Section~\ref{subsec:trdata}, not all light curves have nonzero flux values at each of the 117 data points. Naively computing the loss for these light curves gives a numerical error as the $\log_{10} 0 = -\infty$. Removing these light curves entirely from the training data would prohibit the NNs from learning the parts of the light curves which do have calculated flux values, which is not ideal. Thus, we modified the loss function such that it disregards these missing values. The expression for the loss function for a batch becomes:

\begin{equation}
\text{MAE}(\textbf{y}{\text{true}}, \textbf{y}{\text{pred}}) = \frac{1}{N} \sum_{i=1}^{B} \sum_{j=1}^{T} w_{ij} |y_{ij, \text{true}} - y_{ij, \text{pred}}|
\end{equation}

\begin{equation}
w_{ij} =
\begin{cases}
0, & \text{if}\ y_{ij, \text{true}} = 0 \\
1, & \text{otherwise}
\end{cases}
\end{equation}

\begin{equation}
N = \sum_{i=1}^{B} \sum_{j=1}^{T} w_{ij}
\end{equation}

Here, $\textbf{y}{\text{true}}$ and $\textbf{y}{\text{pred}}$ represent the true and predicted flux values, respectively. $B$ is the batch size (128), and $T$ is the number of time bins (117). The index $i$ iterates over the light curves in the batch, and the index $j$ iterates over the time bins. The variable $w_{ij}$ is a weight that determines whether a specific element contributes to the loss function, and $N$ is the total number of non-zero elements in the batch.

The NNs then only train on parts of the light curves with nonzero flux values. Importantly, this does not mean that the NNs learn to output zero at the zero flux values in the training data. Instead, they will learn to extrapolate the flux to these regions based on the light curves that do have complete coverage. This method is perhaps less physically motivated than, for example, using closure relations (see e.g,~\citealt{gao_complete_2013}) to fill in the zero flux gaps. It does rely only on the \texttt{BOXFIT} calculations, however, and is straightforward to implement in the NN training procedure. 

\section{DEEPGLOW RESULTS}
\label{sec:DGresults}
\begin{figure*}[hbt!]
    \centering
    \includegraphics[width=0.48\textwidth]{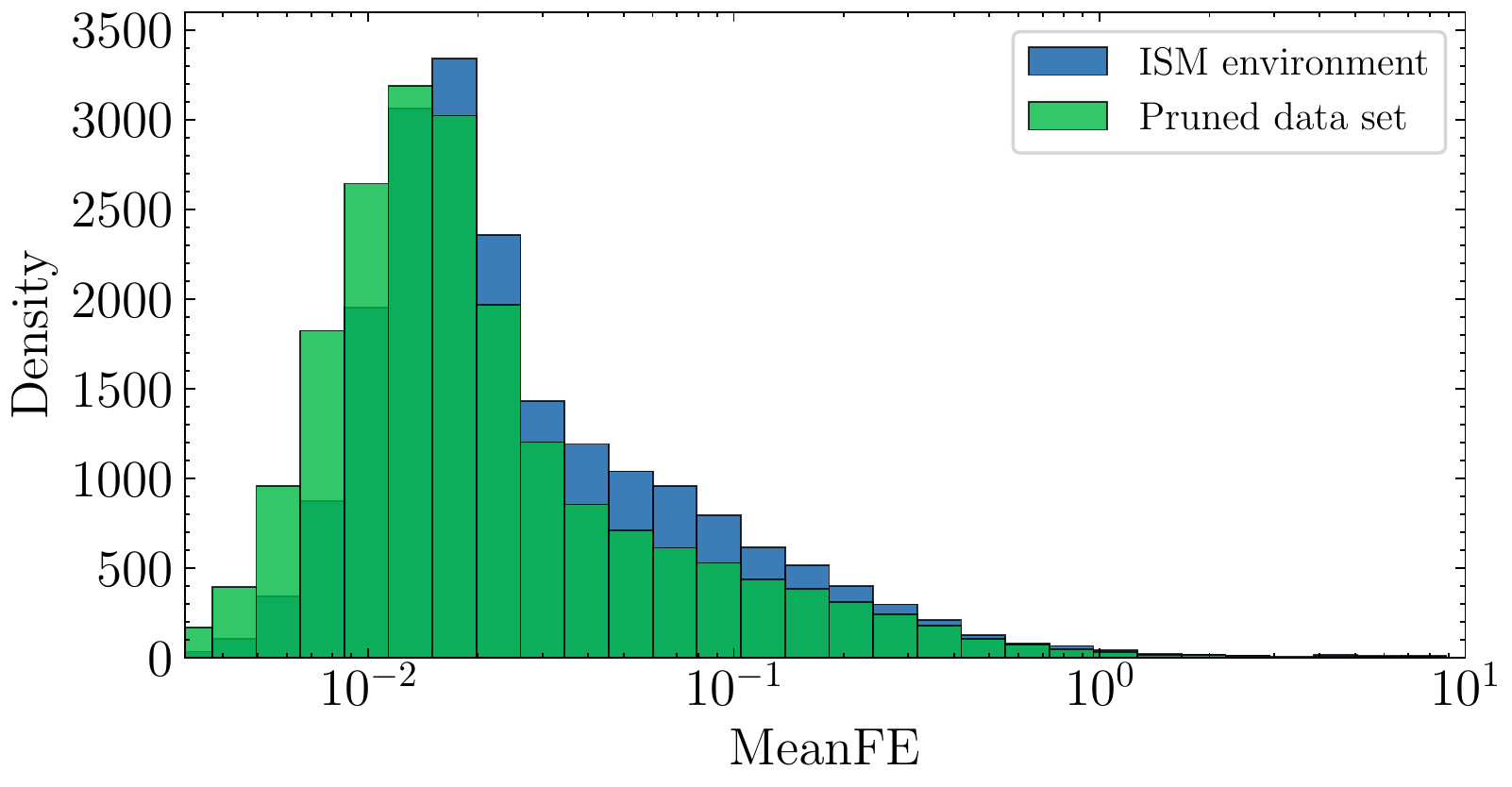}
    \includegraphics[width=0.48\textwidth]{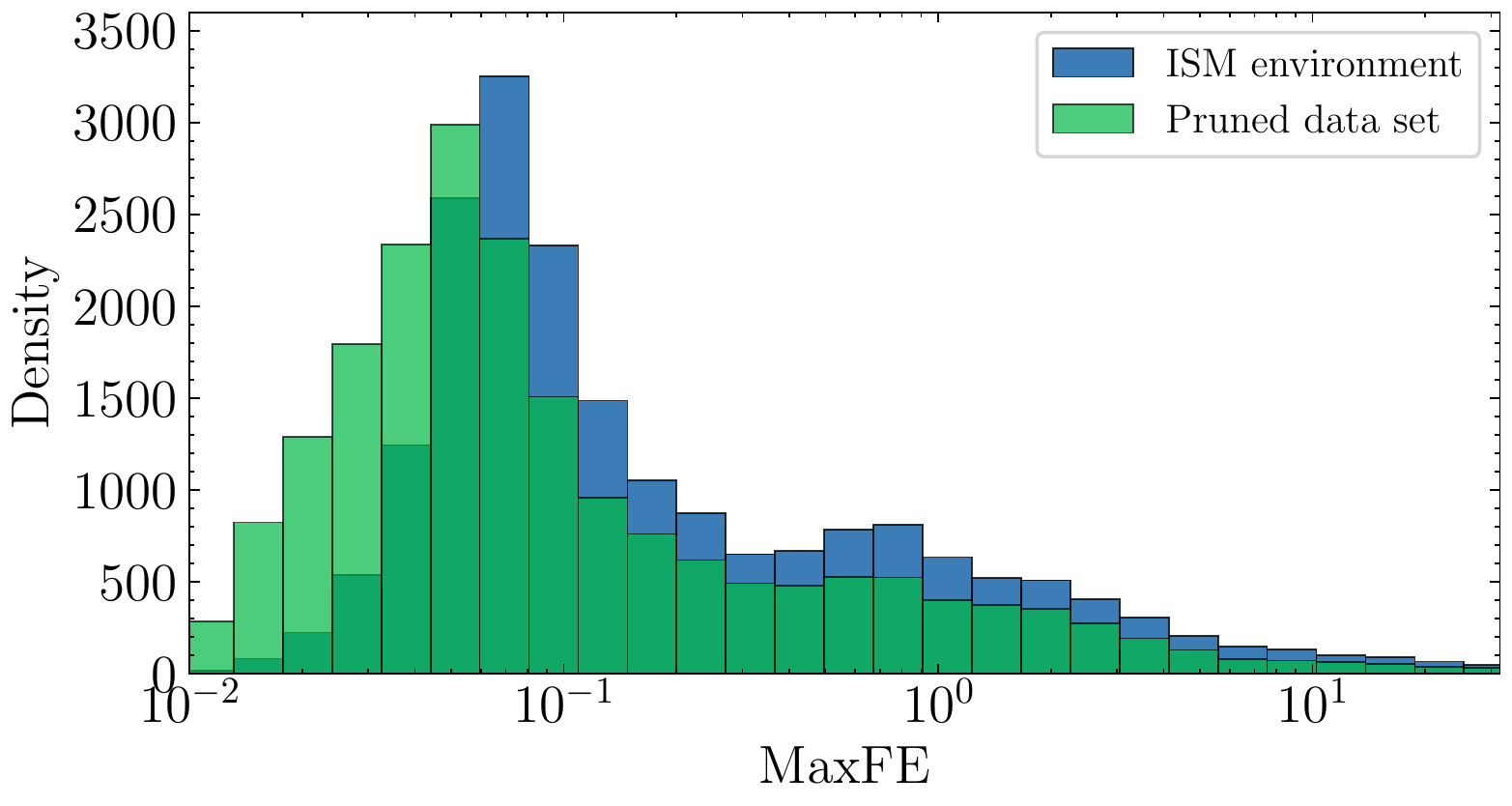}
    \includegraphics[width=0.48\textwidth]{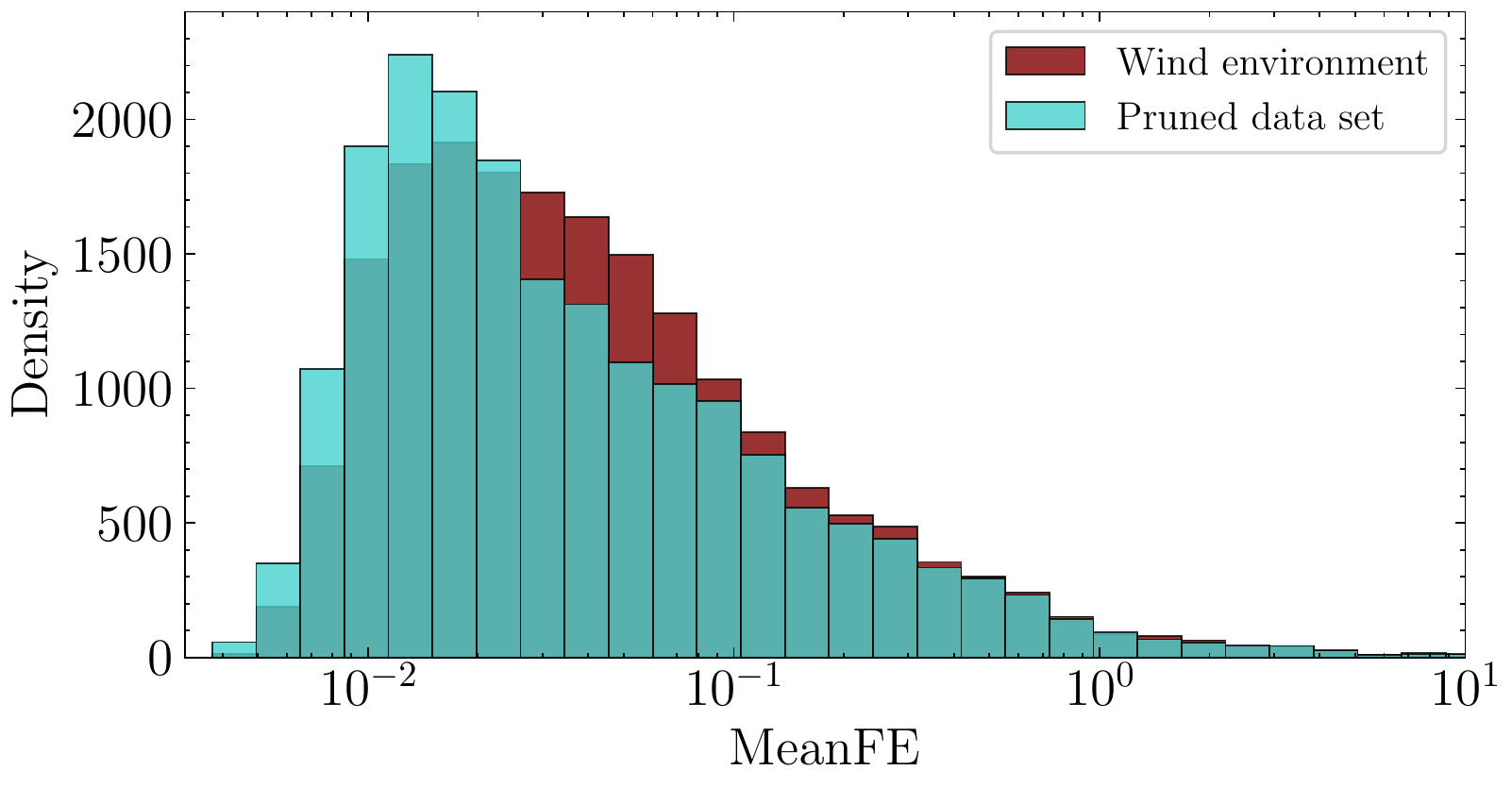}
    \includegraphics[width=0.48\textwidth]{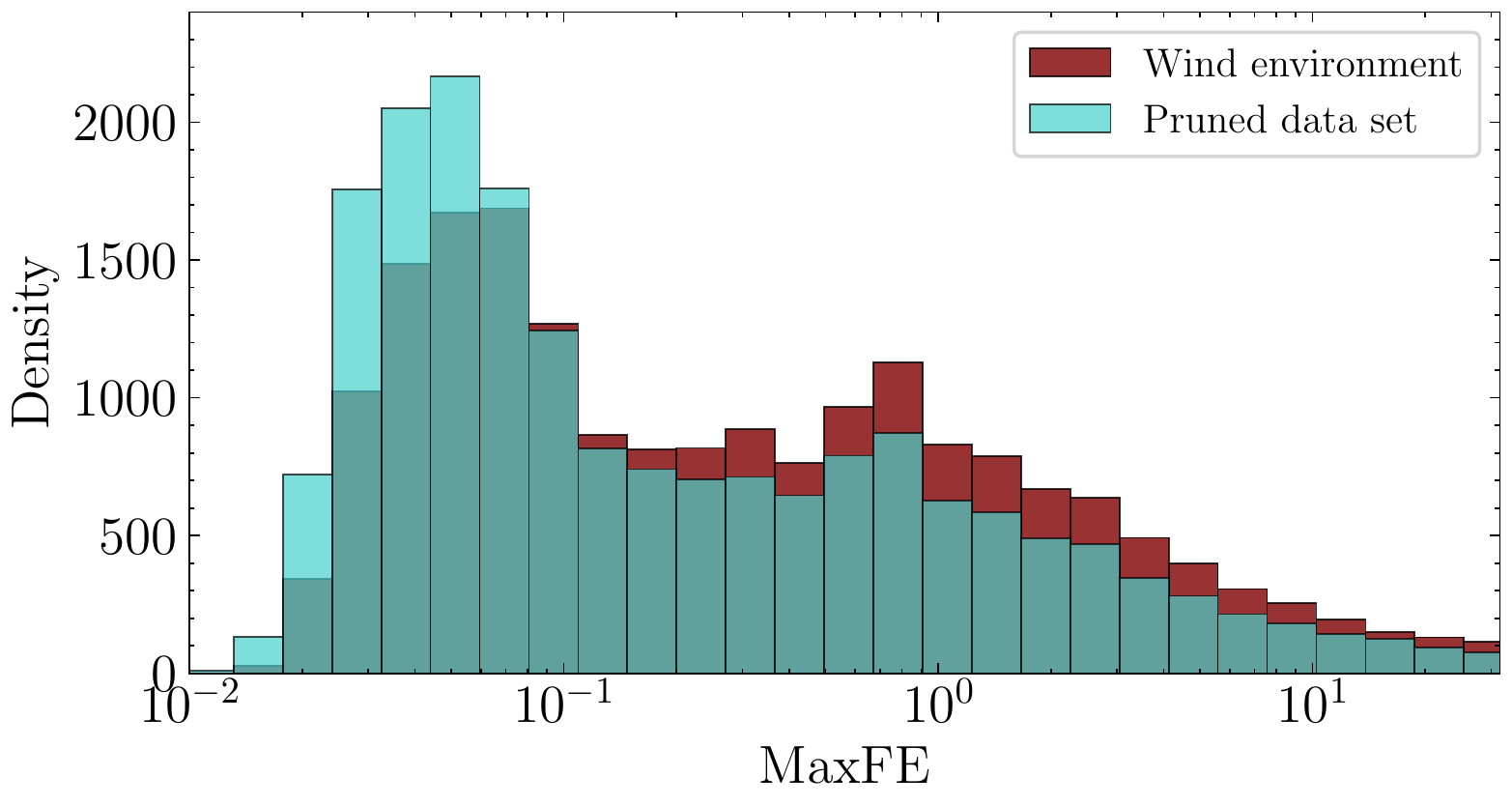}
    \caption{MeanFE and MaxFE distributions scaled logarithmically on the horizontal axis of the light curves in our test dataset. The upper two panels show these distributions assuming an ISM progenitor environment while a wind progenitor environment is used in the lower two panels. In each panel, the pruned dataset refers to the distributions calculated over a smaller observer time range starting from $t \ \approx \ 1$ day.}
    \label{fig:FE_dist}
\end{figure*}

During the training stage of the NNs we calculated the median fractional error over all flux values in the test dataset at the same time to gauge the accuracy of \texttt{DeepGlow}. To get a more complete picture of how well it performs in a production setting, we followed a similar approach to~\cite{kerzendorf_dalek_2021} and defined for each light curve in our test dataset:
\begin{align}
    \mathrm{MeanFE} &= \frac{1}{N} \sum_{i=1}^{N} \frac{|d^{DG}_{k,i}-d^{BF}_{k,i}|}{d^{BF}_{k,i}}, \\
    \mathrm{MaxFE} &= \mathrm{Max}_{i=1}^{N} \frac{|d^{DG}_{k,i}-d^{BF}_{k,i}|}{d^{BF}_{k,i}},
\end{align}
with $N \ (\ = \ 117)$ the amount of data points, and $d_{k,i}$ the flux value of the $i$-th data point of the $k$-th light curve in our test dataset generated by either \texttt{DeepGlow} or \texttt{BOXFIT}.

\begin{figure*}[t]
    \centering
    \includegraphics[width=0.49\textwidth]{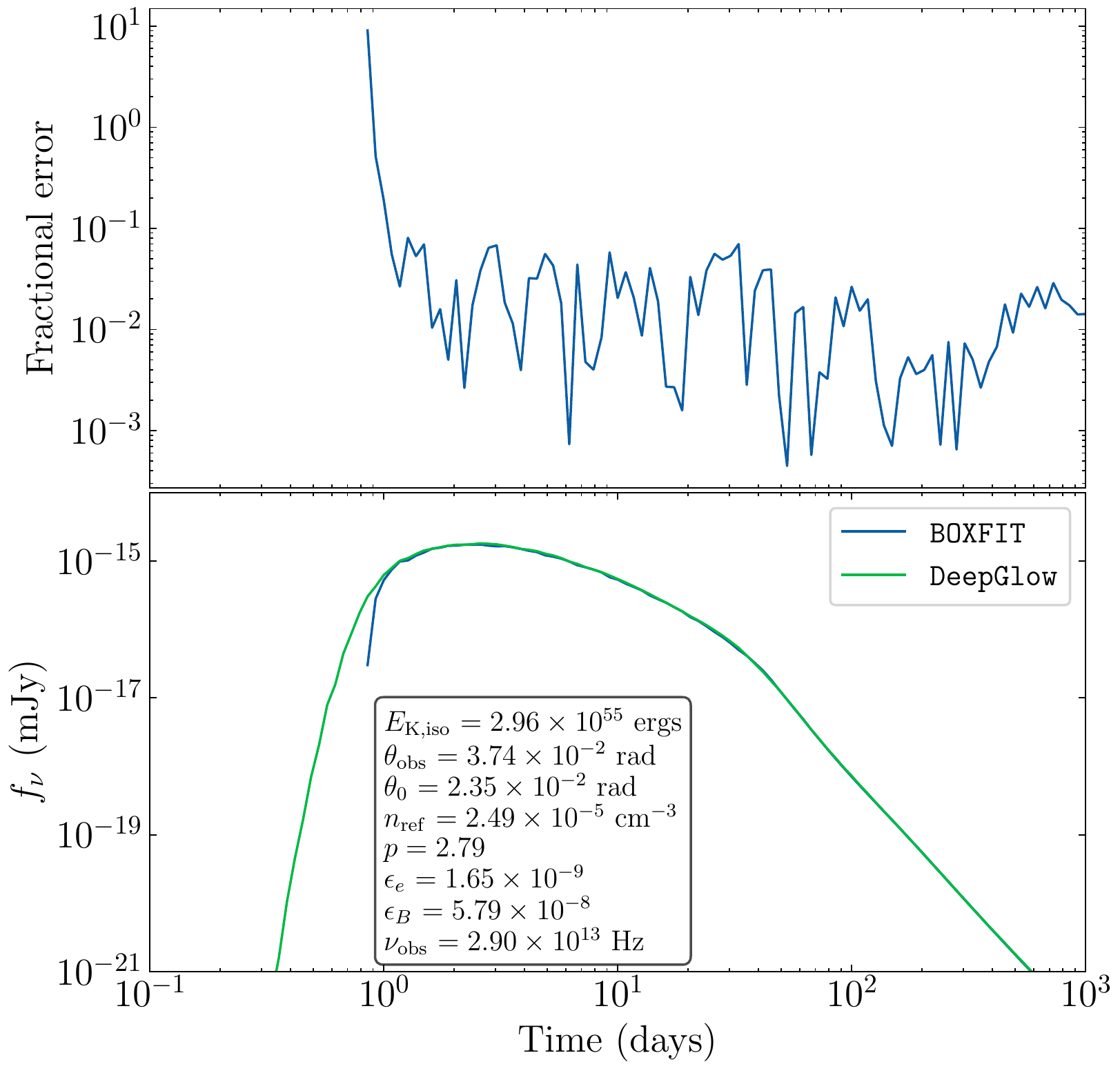}
    \includegraphics[width=0.49\textwidth]{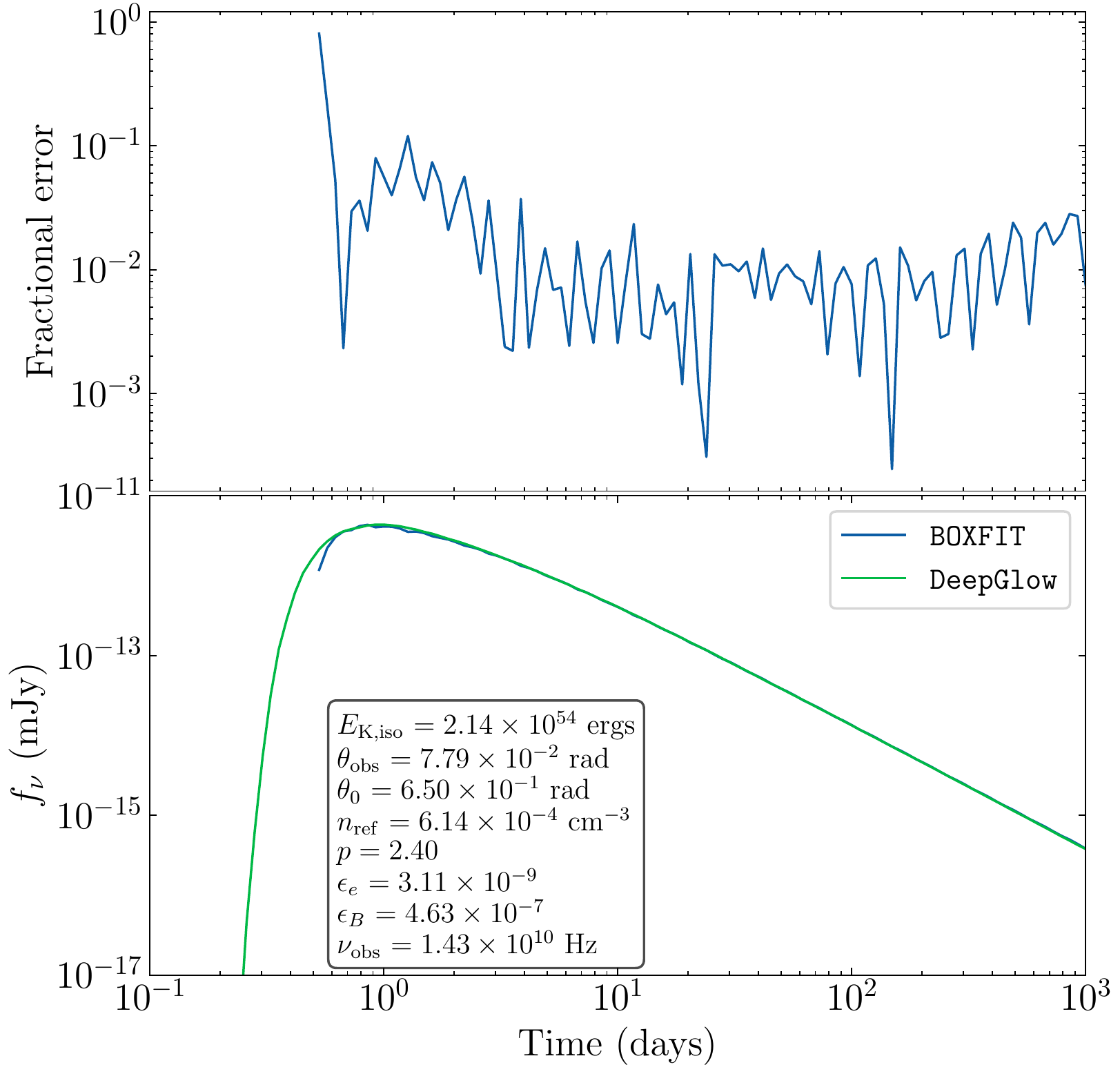}
    \caption{Examples of light curves in the test data sets where the $\mathrm{MaxFE}$ is concentrated at the edge of the observer times which are still covered by \texttt{BOXFIT} calculations. The left panel shows an example for the ISM environment whereas the right panel shows an example for the wind environment. In both panels, the top plot shows the fractional error of the \texttt{DeepGlow} prediction versus the \texttt{BOXFIT} calculations, whereas the bottom plot shows both the \texttt{BOXFIT} (blue) and \texttt{DeepGlow} (green) light curves. The inset shows the corresponding parameter values.}
    \label{fig:lc_examples}
\end{figure*}

For each progenitor environment, we calculated the $\mathrm{MeanFE}$ and $\mathrm{MaxFE}$ distributions over the test dataset for the full observer time range. In addition, we also calculated these statistics assuming a more limited range starting from $t_\mathrm{obs} \ \approx \ 1$ day ($i \ = \ 30$), after which most GRB afterglow observations usually take place. We will refer to this second case as the pruned dataset which has the same amount of light curves as the test dataset but takes fewer flux values into consideration. Both distributions, the $\mathrm{MeanFE}$ and $\mathrm{MaxFE}$ statistics calculated over the full test dataset and the pruned dataset, are shown in Figure~\ref{fig:FE_dist}, scaled logarithmically on the horizontal axis. The median of each $\mathrm{MeanFE}$ and $\mathrm{MaxFE}$ distribution shown is given in Table~\ref{tab:medvalues}.

\begin{table}[t]
\begin{tabular}{|l|l|l|} 
\toprule
Median values                    & ISM environment & Wind environment \\ \hline
$\mathrm{MeanFE}$                & 0.020           & 0.034            \\
$\mathrm{MeanFE}$ pruned dataset & 0.016           & 0.026            \\
$\mathrm{MaxFE}$                 & 0.104           & 0.228            \\
$\mathrm{MaxFE}$ pruned dataset  & 0.062           & 0.099        
\\ \bottomrule
\end{tabular}
\caption{Median values of the MeanFE and MaxFE distributions for each progenitor environment.}\label{tab:medvalues}
\end{table}

\begin{figure*}[hbt!]
    \centering
    \includegraphics[width=0.9\textwidth]{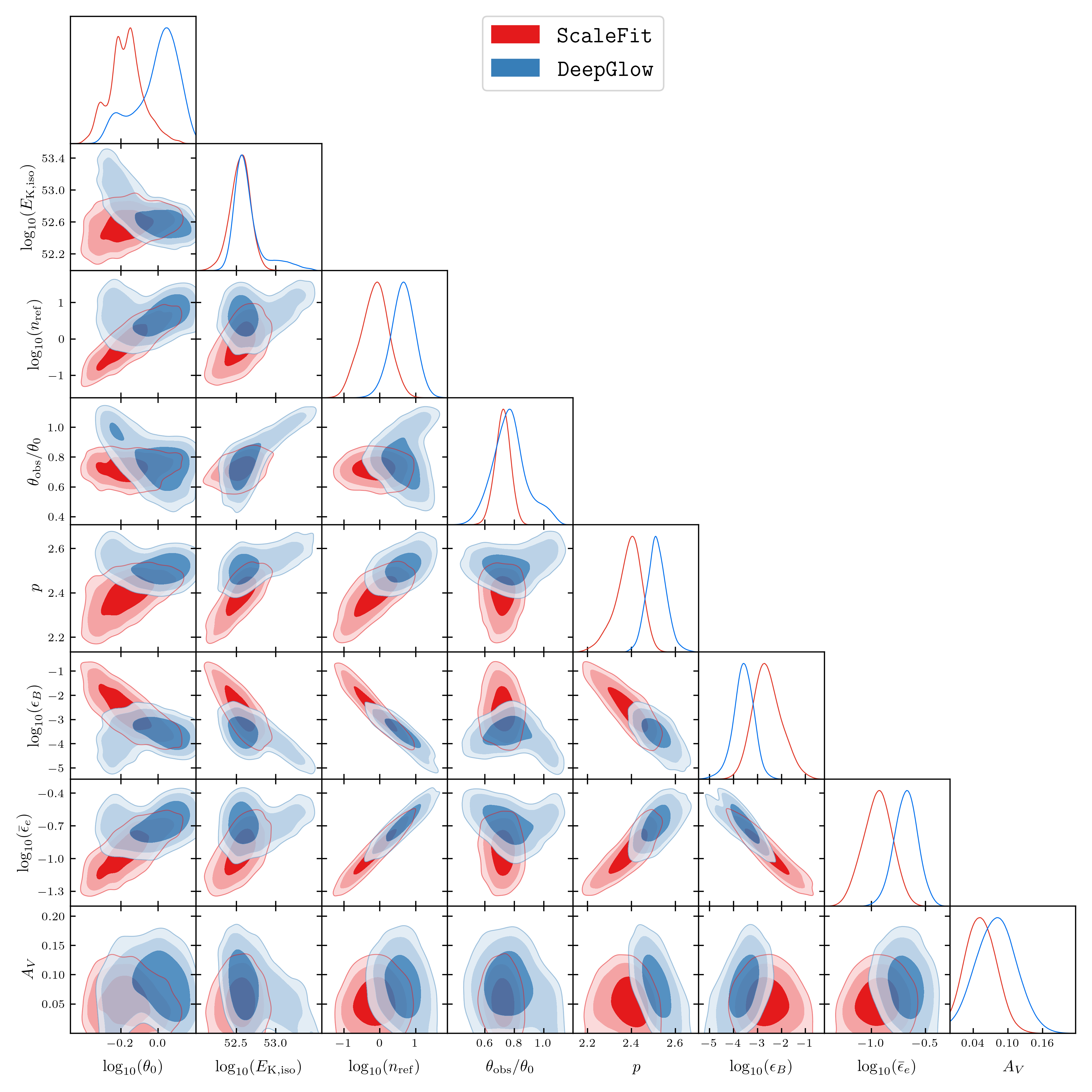}
    \caption{Posterior distribution of the GRB afterglow parameters for GRB970508 assuming an ISM progenitor environment. The likelihood was calculated using either the \texttt{ScaleFit} model in red or our \texttt{DeepGlow} model in blue.}
    \label{fig:marg_dist_ism}
\end{figure*}

In general, the average NN reproduction error per light curve, i.e. $\mathrm{MeanFE}$, is a few percent which is well below the typical measurement error on GRB afterglow observations. For the ISM environment, the $\mathrm{MaxFE}$ on most generated light curves is smaller than the typical measurement error as well, certainly when looking at the pruned dataset. The $\mathrm{MaxFE}$ in the wind environment case can become quite high ($>30\%$) but this improves considerably in the pruned dataset. The better performance in the pruned dataset can be traced back to better \texttt{BOXFIT} calculation coverage for later observer times, see Fig.~\ref{fig:percentage_zeroflux}. There are fewer data points to train on at $t_\mathrm{obs} < 1$ days, and the flux evolution is more sporadic, making it harder to predict. We observe that the average $\mathrm{MaxFE}$ is substantially higher for light curves where the \texttt{BOXFIT} coverage is incomplete. At the limits of the observer times which still have nonzero fluxes for these light curves, the simulated \texttt{BOXFIT} flux can drop off very rapidly and/or have large simulation noise. As mentioned in Section~\ref{subsec:trdata}, this is also a sign of incomplete simulation coverage even though the flux is still nonzero. Slight errors in e.g. the reproduced slope by \texttt{DeepGlow} can easily lead to large fractional errors compared to \texttt{BOXFIT} in these instances. We show two examples of this for light curves in our test dataset in Fig.~\ref{fig:lc_examples}. Thus, large fractional errors and data points with zero fluxes are usually in close proximity with one another with respect to their observer times. Consequently, $\mathrm{maxFE}$ errors are often found in regions where \texttt{BOXFIT} calculations are intrinsically inaccurate and NN emulation is not meaningful anyways. Because the \texttt{BOXFIT} coverage is less of an issue in the pruned dataset, the median $\mathrm{MaxFE}$ will also be lower.

Incomplete \texttt{BOXFIT} coverage is more likely in quite extreme regions of the GRB afterglow parameter space (e.g. very high energies in combination with very low densities) which may not be where most observed GRB afterglows reside. Still, for the wind environment in particular, it is possible that the reproduction error will lead to a significant systematic error contribution compared to \texttt{BOXFIT} when fitting some observed data points. Moreover, while \texttt{DeepGlow} is trained on light curves with a fixed observer time grid, we use linear interpolation to allow for arbitrary observer times within the limits $t_0$ and $t_1$. This could also be an extra source of systematics not captured in the $\mathrm{MeanFE}$ and $\mathrm{MaxFE}$ distributions.

For many GRB afterglow datasets, we are confident that \texttt{DeepGlow} can be used to fit the data in place of \texttt{BOXFIT} with good accuracy. We advise caution, however, when interpreting GRB afterglow datasets with best fit parameter values which lie in regions of the parameter space where \texttt{BOXFIT} has incomplete coverage over the observation times of the measurements.

The mean and standard deviation of the \texttt{DeepGlow} evaluation time are just $2.2 \pm 0.2$ ms on a single thread of our computing cluster. In contrast, the \texttt{BOXFIT} mean and standard deviation evaluation time on 40 threads of a single node are $29.6 \pm 4.6$ s. \texttt{DeepGlow} thus represents an approximate $10^4$ factor speedup in evaluation, which further increases if less threads for \texttt{BOXFIT} parallel execution are available, making parameter estimation with the physics of \texttt{BOXFIT} possible.

In the next section, we will use \texttt{DeepGlow} to estimate the parameters of the GRB970508 afterglow. In~\ref{app:extraNNacc}, we provide some additional figures related to the training of the NNs.

\section{TEST CASE: GRB970508}
\label{sec:testcase}
\subsection{Gaussian Process framework}
\label{subsec:GPF}
\begin{figure*}[hbt!]
    \centering
    \includegraphics[width=0.9\textwidth]{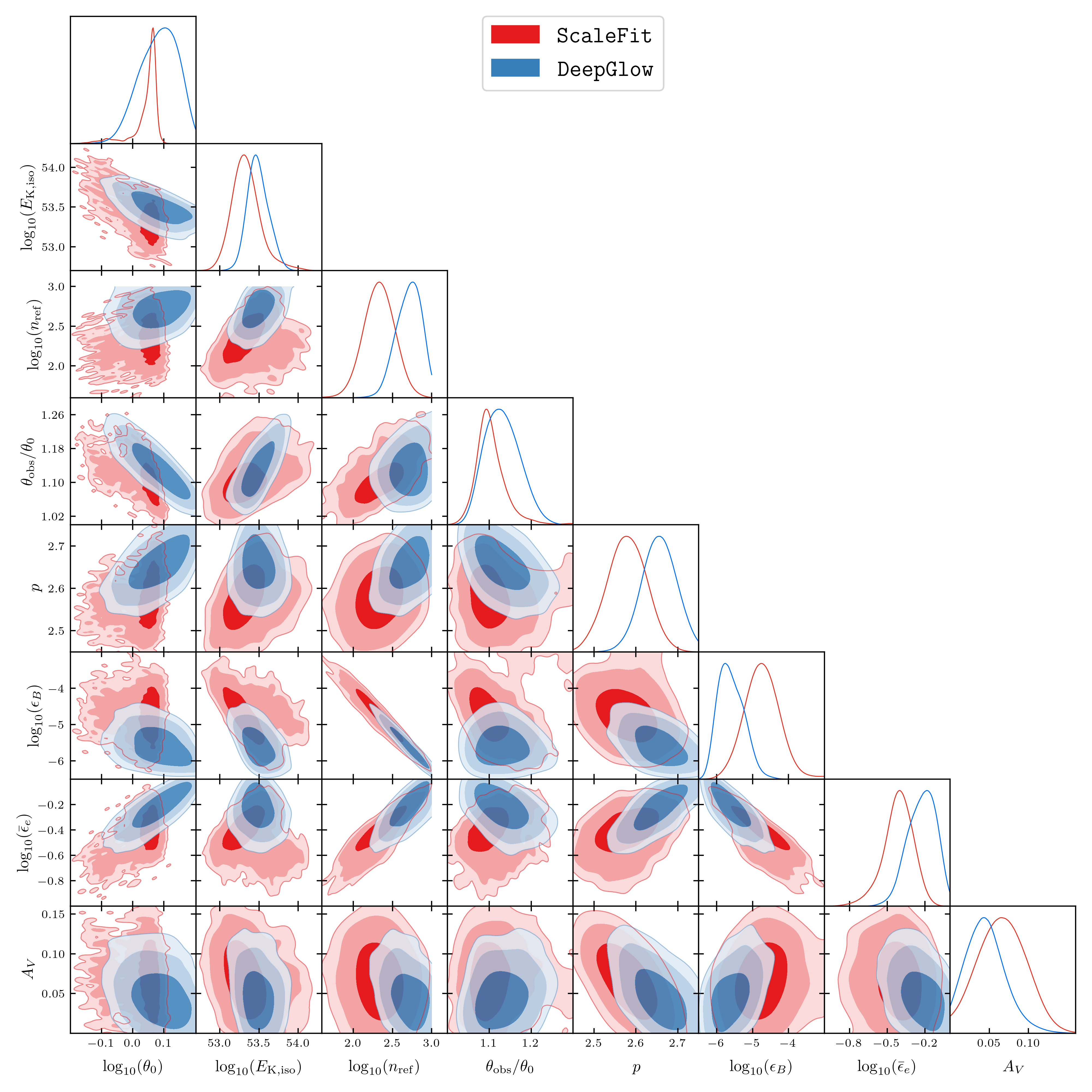}
    \caption{Same as Figure~\ref{fig:marg_dist_ism}, except now assuming a wind progenitor environment.}
    \label{fig:marg_dist_wind}
\end{figure*}

As a first scientific application of \texttt{DeepGlow}, we inferred the properties of the afterglow of GRB970508 using the methods in\defcitealias{aksulu_exploring_2022}{A22} \citet[][hereafter referred to as  \citetalias{aksulu_exploring_2022}]{aksulu_exploring_2022}. They build on the methods in~\citet{aksulu_new_2020} and use a Gaussian Process (GP,~\citealt{rasmussen_gaussian_2006}) framework to estimate the parameters of GRB afterglow datasets while allowing for unknown systematics to be modeled simultaneously as well. Especially when considering unmodeled systematic effects such as scintillation at radio frequencies, conventional $\chi^2$ fitting methods can lead to underestimated uncertainties on parameters~\citep{aksulu_new_2020}. By modeling the systematics using GPs, we have a much more robust method to obtain parameter estimates. In \Aks, the GRB afterglow model of choice is \texttt{ScaleFit} (\citealt{ryan_gamma-ray_2015}, Ryan et al. in prep). It is a semi-analytical model which uses pre-calculated spectral tables from \texttt{BOXFIT} to model the GRB afterglow in different spectral regimes. \texttt{ScaleFit} is a computationally inexpensive alternative to \texttt{BOXFIT} and, also in contrast to \texttt{BOXFIT}, is valid in all spectral regimes (see \Aks~for details). A downside to \texttt{ScaleFit} is the assumptions it has to make about the sharpness of spectra around break frequencies. This follows naturally from the radiative transfer approach that \texttt{BOXFIT} uses and is thus incorporated in \texttt{DeepGlow} too. The evaluation time for \texttt{ScaleFit}, after generating the necessary spectral tables, is 0.9 $\pm$ 0.1 ms on our computing cluster. Taking into account the sampling overhead, parameter estimation runs are similar in runtime to those with \texttt{DeepGlow}.

\Aks~use the \texttt{MultiNest} nested sampler~\citep{feroz_multinest_2009} through the Python implementation \texttt{PyMultiNest}~\citep{buchner_x-ray_2014} to sample the GP likelihood~(Eq. 2 of~\citealt{aksulu_new_2020}). This nested sampling approach requires on the order of 100 000 likelihood evaluations for each fit. Because \texttt{BOXFIT} can generate a light curve for only one frequency at a time, and the GRB970508 observations cover twelve different frequencies, \texttt{BOXFIT} would have to be run twelve times for one likelihood evaluation. This becomes intractable to calculate, certainly on a repeated basis when, e.g., fitting an afterglow dataset multiple times with different \texttt{MultiNest} settings. We thus compared the results using \texttt{DeepGlow} to the results of \Aks~using \texttt{ScaleFit}. While this is not a direct comparison between \texttt{DeepGlow} and \texttt{BOXFIT}, \texttt{ScaleFit} is similar enough to \texttt{BOXFIT} to give a general indication of how well \texttt{DeepGlow} works in practice. We performed two fits per afterglow model, one for each progenitor environment.

An important difference between \texttt{DeepGlow} and \texttt{ScaleFit} lies in how the model parameters $p$ and $\epsilon_e$ are handled. \texttt{ScaleFit} fits the parameter $\Bar{\epsilon}_e \equiv \frac{p-2}{p-1} \epsilon_e$ instead of $\epsilon_e$ to allow for fits with $p \ < \ 2$. For the sake of comparison, we also fitted $\Bar{\epsilon}_e$ and calculated $\epsilon_e$ from $\Bar{\epsilon}_e$. \Aks~extended the prior range of $p$ below two as well. This is not possible for \texttt{DeepGlow} as \texttt{BOXFIT}, on which it is trained, cannot calculate light curves or spectra with $p \ < \ 2$. Here we are thus limited to fits with $p \ > \ 2$ which might hamper any comparisons. The posteriors of GRB970508 using \texttt{ScaleFit} have little support for $p \ < \ 2$, however, and we assume that our more limited prior range did not influence the results much in this case (though it might for other GRB afterglows which do have support for $p \ < \ 2$, see~\Aks). The rest of the methodology, including the prior ranges and \texttt{MultiNest} settings, was the same as those in \Aks. As such, we also included $A_{V}$, the rest-frame value for host galaxy dust extinction, as a free parameter in the model.

We will limit ourselves to quantitative comparisons between the two afterglow models in the next section. An in-depth comparison of the differences in estimated parameters that may arise in relation to any physical differences between \texttt{DeepGlow}, i.e. \texttt{BOXFIT}, and \texttt{ScaleFit} is beyond the scope of this work.    
\subsection{Results}
\label{subsec:GRBres}
\begin{figure*}[hbt!]
    \centering
    \includegraphics[width=0.49\textwidth]{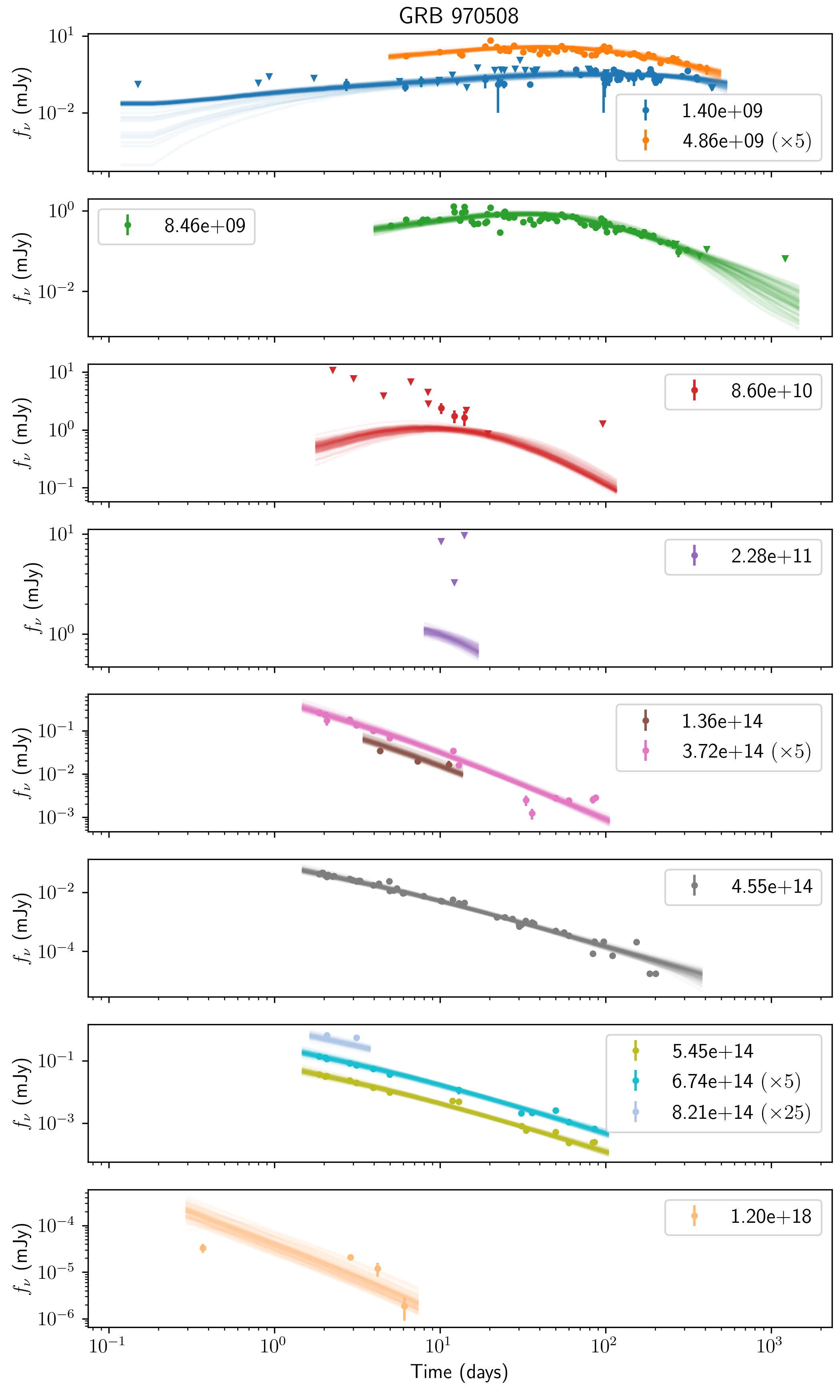}
    \includegraphics[width=0.49\textwidth]{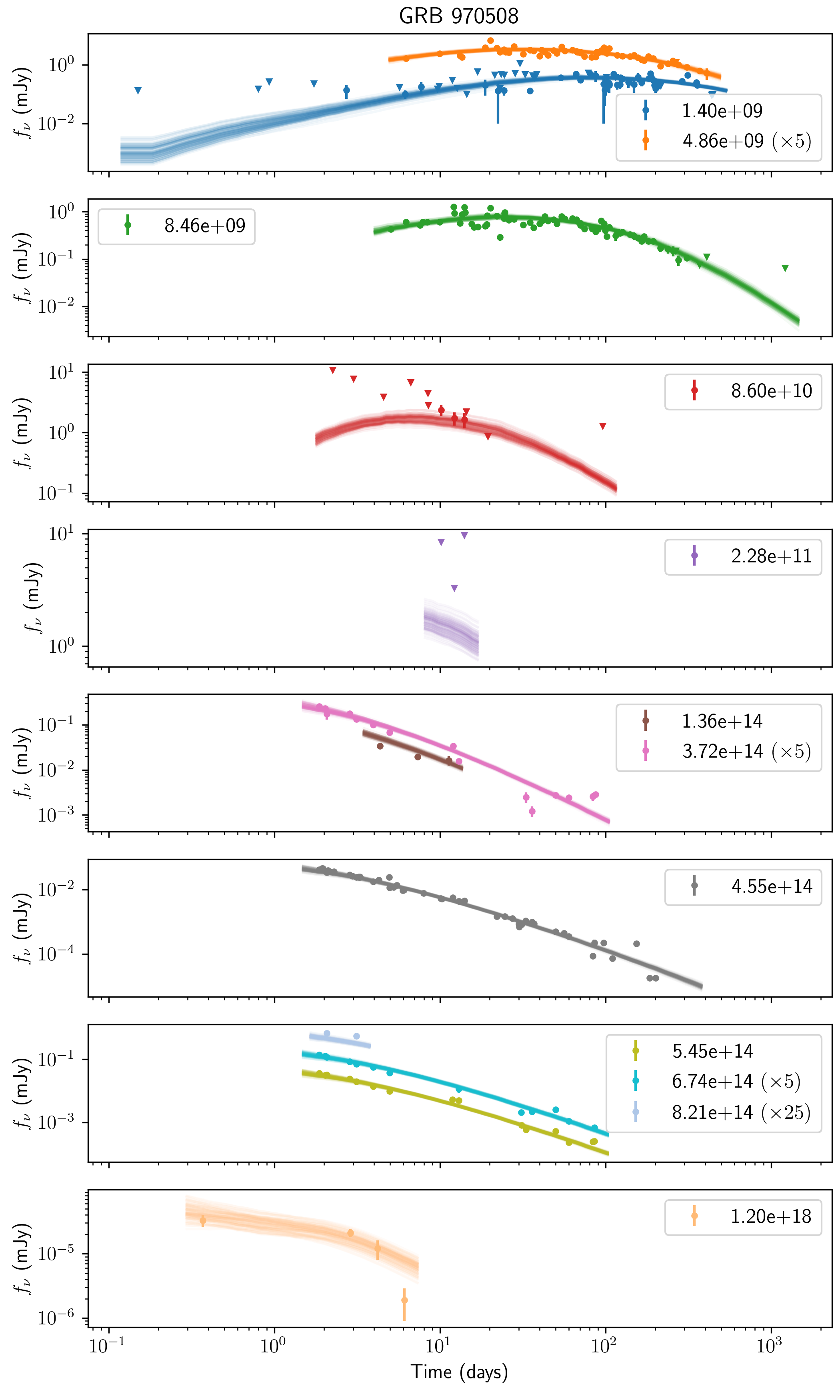}
    \caption{Fit results using \texttt{DeepGlow} for the dataset of GRB970508. The panels on the left side of the figure show the fit assuming an ISM progenitor environment while the panels on the right show the fit for a wind progenitor environment. The colored points correspond to the observed flux densities from the X-ray to radio bands. The legends of each subpanel display the observer frequency in Hz. The multiplication factor for a legend item indicates that the flux density is multiplied by this factor. The dots represent actual measurements while triangles correspond to the 3-$\sigma$ upper limits. From the posteriors of Fig.~\ref{fig:marg_dist_ism} and Fig.~\ref{fig:marg_dist_wind}, respectively, 100 parameter sets are drawn randomly and the \texttt{DeepGlow} light curves are computed and shown as semi-transparent solid lines. More opaque regions thus correspond to higher posterior probabilities.}
    \label{fig:posterior_lc}
\end{figure*}

\begin{table*}
\RawFloats
\begin{minipage}{0.48\columnwidth}
\centering
\caption{Median values of the marginal posterior distributions for the afterglow parameters of GRB970508 assuming an ISM environment. Quoted uncertainties are at the 68\% level. A match indicates if the uncertainty intervals overlap for the two afterglow models.}
\label{tab:margISM}
\begin{tabular} {| l | l | l | l |}
\hline
Parameter & \texttt{DeepGlow} & \texttt{ScaleFit} & Match \\
\hline
{$\log_{10}\theta_0$} & $0.02^{+0.13}_{-0.08}$ & $-0.17^{+0.09}_{-0.09}$ & $ \quad \times$ \\
{$\log_{10}E_{K, \mathrm{iso}}$} & $52.60^{+0.11}_{-0.15}$ & $52.56^{+0.13}_{-0.12}$ & $ \quad \checkmark$ \\
{$\log_{10}n_{\mathrm{ref}}$} & $0.66^{+0.33}_{-0.33}$ & $-0.12^{+0.40}_{-0.39}$ & $ \quad \times$ \\
{$\theta_{\mathrm{obs}} / \theta_0$} & $0.77^{+0.10}_{-0.12}$ & $0.72^{+0.05}_{-0.05}$ & $ \quad \checkmark$ \\
{$p$} & $2.51^{+0.04}_{-0.04}$ & $2.39^{+0.06}_{-0.05}$ & $ \quad \times$ \\
{$\log_{10}\epsilon_B$} & $-3.58^{+0.44}_{-0.39}$ & $-2.64^{+0.58}_{-0.63}$ & $ \quad \checkmark$ \\
{$\log_{10}\bar{\epsilon}_e$} & $-0.68^{+0.11}_{-0.11}$ & $-0.94^{+0.14}_{-0.14}$ & $ \quad \times$ \\
{$A_V$} & $0.08^{+0.03}_{-0.04}$ & $0.05^{+0.03}_{-0.03}$ & $ \quad \checkmark$ \\
\hline
\end{tabular}
\end{minipage}\hfill
\begin{minipage}{0.48\columnwidth}
\centering
\caption{Same as Table~\ref{tab:margISM}, except now assuming a wind progenitor environment.}
\label{tab:margwind}
\begin{tabular} {| l | l | l | l |}
\hline
Parameter & \texttt{DeepGlow} & \texttt{ScaleFit} & Match \\
\hline
{$\log_{10}\theta_0$} & $0.09^{+0.07}_{-0.06}$ & $0.06^{+0.02}_{-0.02}$ & $ \quad \checkmark$ \\
{$\log_{10}E_{K, \mathrm{iso}}$} & $53.48^{+0.13}_{-0.14}$ & $53.33^{+0.16}_{-0.20}$ & $ \quad \checkmark$ \\
{$\log_{10}n_{\mathrm{ref}}$} & $2.72^{+0.18}_{-0.14}$ & $2.33^{+0.21}_{-0.20}$ & $ \quad \times$ \\
{$\theta_{\mathrm{obs}} / \theta_0$} & $1.13^{+0.04}_{-0.05}$ & $1.10^{+0.03}_{-0.03}$ & $ \quad \checkmark$ \\
{$p$} & $2.66^{+0.04}_{-0.04}$ & $2.58^{+0.05}_{-0.05}$ & $ \quad \checkmark$ \\
{$\log_{10}\epsilon_B$} & $-5.64^{+0.35}_{-0.38}$ & $-4.73^{+0.48}_{-0.51}$ & $ \quad \times$ \\
{$\log_{10}\bar{\epsilon}_e$} & $-0.21^{+0.12}_{-0.11}$ & $-0.41^{+0.13}_{-0.11}$ & $ \quad \checkmark$ \\
{$A_V$} & $0.04^{+0.02}_{-0.03}$ & $0.07^{+0.03}_{-0.03}$ & $ \quad \checkmark$ \\
\hline
\end{tabular}
\end{minipage}
\end{table*}

We show the posteriors in Figure~\ref{fig:marg_dist_ism} (ISM environment) and Figure~\ref{fig:marg_dist_wind} (wind environment). The median values and 68\% credible intervals of each parameter, and if these values overlap for \texttt{DeepGlow} and \texttt{ScaleFit}, are given in Table~\ref{tab:margISM} (ISM environment) and Table~\ref{tab:margwind} (wind environment). Note that the values presented here for \texttt{ScaleFit} are somewhat different than those in \Aks~because we have calculated, for simplicity, the median and not the mode of the distributions. Furthermore, small sampling differences between parameter estimation runs can arise as well.

For the most part, the parameter estimates of \texttt{DeepGlow} and \texttt{ScaleFit} overlap fairly well. For both environments, the estimates for $E_\mathrm{K,iso}$ and $\theta_\mathrm{obs}/\theta_0$ in particular are in good agreement between the two afterglow models. An exact match for all parameters is not expected in any case, as \texttt{ScaleFit} does have some distinct differences over \texttt{BOXFIT}, as mentioned. For the marginalized distributions that do not overlap within 1$\sigma$, there is usually only a slight discrepancy. Furthermore, strong correlations between parameters, e.g. $n_\mathrm{ref}$ and $\epsilon_B$, are captured well by \texttt{DeepGlow} in accordance with \texttt{ScaleFit}. 

A big benefit of nested sampling is the direct computation of the Bayesian evidence $\mathcal{Z}$ as part of the sampling procedure. This allows us to give both a preference in terms of the progenitor environment as well as the afterglow model by calculating the ratio of evidence values, i.e. the Bayes factor $\mathrm{BF}$. We follow~\citet{kass_bayes_1995} for interpretation. 

We find a decisive preference for the wind environment over the ISM environment for both \texttt{DeepGlow} and \texttt{ScaleFit} with $\mathrm{BF}_\mathrm{wind/ISM} \sim 10^4$ and $\mathrm{BF}_\mathrm{wind/ISM} \sim 10^3$, respectively, similar to what was found in \Aks. We also find a strong preference for \texttt{DeepGlow} over \texttt{ScaleFit} in the ISM case with $\mathrm{BF}_\mathrm{DG/SF} \sim 20$ and a decisive preference in the wind case with $\mathrm{BF}_\mathrm{DG/SF} \sim 10^3$.

An important caveat to the parameter estimates of \texttt{DeepGlow} for the wind environment, is that for most values of $\bar{\epsilon}_e$ and $p$ in the posterior, $\epsilon_e$ becomes greater than one which is unphysical. To a lesser degree, this is true for the posterior of \texttt{ScaleFit} (wind environment) as well. In these instances, instead of setting $\xi_N = 1$, a lower value is perhaps better suited, e.g. $\xi_N = 0.1$, to scale down the other degenerate parameters ($E_\mathrm{K,iso}$, $n_\mathrm{ref}$, $\bar{\epsilon}_e$, and $\epsilon_B$) to more physical values~\citep{eichler_efficiency_2005}. As in most literature, including \Aks~though not in~\citet{aksulu_new_2020}, $\xi_N$ is fixed canonically to 1, we did not use another value in our study here.

In Fig.~\ref{fig:posterior_lc}, we plot the broadband GRB970508 dataset we fitted using \texttt{DeepGlow}. We drew 100 parameter sets randomly from the posteriors in Fig.~\ref{fig:marg_dist_ism} and Fig.~\ref{fig:marg_dist_wind} and drew the resulting light curve for each set using \texttt{DeepGlow}. This gives a visual confirmation that the fits are of good quality for both progenitor environments.

Overall, the results for \texttt{DeepGlow} and \texttt{ScaleFit} seem consistent. While additional systematics by \texttt{DeepGlow} compared to \texttt{BOXFIT} could change the results slightly, see the next section, we contend that a direct fit with \texttt{BOXFIT} would produce similar results to \texttt{DeepGlow}. Characterizing other GRB afterglow datasets with \texttt{DeepGlow} could thus provide an interesting avenue for a more thorough comparison between the physics of \texttt{BOXFIT} and \texttt{ScaleFit}.
\subsection{Systematics}
To characterize the systematics of \texttt{DeepGlow}, we reran the fit on GRB970508 assuming a wind environment. We used a different NN realisation this time from the same training run but trained for 1940 epochs. It has a very similar median error calculated over all data points in the test dataset compared to the primary NN realisation used which was trained for 1900 epochs, see~\ref{app:extraNNacc}. Because the MaxFE on some reproduced light curves can become quite large, we may expect significant variation for some data points in the light curves generated between two NN realisations with slightly different weights. Any resulting change in the parameter estimates will give an indication on the influence of the systematic reproduction errors in \texttt{DeepGlow}. We show the results in Table~\ref{tab:DGvsDG}.

The estimates are close and readily within the $1\sigma$ uncertainties for all parameters. Still, even though the overall error is much below the typical measurement error for both NN realisations, there are some variations in the estimated parameters. These are larger than any sampling differences we observed for \texttt{DeepGlow} runs. We attribute these variations to the large MaxFE for certain light curves. While this is not a substantial source of systematic errors, it is something to be taken into consideration.

\begin{table}
\begin{tabular} {| l | l | l | l |}
\hline
Parameter & \texttt{DeepGlow} & \texttt{Alt.~NN} & Match \\
\hline
{$\log_{10}\theta_0$} & $0.09^{+0.07}_{-0.06}$ & $0.10^{+0.06}_{-0.05}$ & $ \quad \checkmark$ \\
{$\log_{10}E_{K, \mathrm{iso}}$} & $53.48^{+0.13}_{-0.14}$ & $53.46^{+0.11}_{-0.13}$ & $ \quad \checkmark$ \\
{$\log_{10}n_{\mathrm{ref}}$} & $2.72^{+0.18}_{-0.14}$ & $2.66^{+0.18}_{-0.15}$ & $ \quad \checkmark$ \\
{$\theta_{\mathrm{obs}} / \theta_0$} & $1.13^{+0.04}_{-0.05}$ & $1.11^{+0.04}_{-0.04}$ & $ \quad \checkmark$ \\
{$p$} & $2.66^{+0.04}_{-0.04}$ & $2.64^{+0.04}_{-0.04}$ & $ \quad \checkmark$ \\
{$\log_{10}\epsilon_B$} & $-5.64^{+0.35}_{-0.38}$ & $-5.46^{+0.35}_{-0.38}$ & $ \quad \checkmark$ \\
{$\log_{10}\bar{\epsilon}_e$} & $-0.21^{+0.12}_{-0.11}$ & $-0.28^{+0.12}_{-0.11}$ & $ \quad \checkmark$ \\
{$A_V$} & $0.04^{+0.02}_{-0.03}$ & $0.05^{+0.02}_{-0.03}$ & $ \quad \checkmark$ \\
\hline
\end{tabular}
\caption{Same as Table~\ref{tab:margwind}, except now comparing two realizations of \texttt{DeepGlow} from the same training run, see text.}\label{tab:DGvsDG}
\end{table}

\section{DISCUSSION AND OUTLOOK}
\label{sec:discussion}
\subsection{DeepGlow use cases}
The current version of \texttt{DeepGlow} emulates the GRB afterglow simulations of \texttt{BOXFIT} with high accuracy. 
The data for the emulator was generated during $\sim$$10^5$ core hours 
and the emulator now produces individual light curves within a few milliseconds.
This efficiency is best put to use for enabling population studies that require a large number of model runs
(e.g., >10$^6$ for \citealt{aksulu_exploring_2022}; $\sim$$10^7$ for \citealt{2022A&A...664A.160B}). 
There, \texttt{DeepGlow} speeds up iterations to run within of order an hour, offering astronomers an interactive exploration approach.

The quick turn-around time may also be beneficial in getting a first estimate of the underlying parameters from  x-ray and optical detections of an afterglow, and inform the radio follow-up campaign. 
Before day 1 there is a $\sim$15\% chance the BOXFIT-based training data was incomplete (Fig.~\ref{fig:percentage_zeroflux}). 
Our light curves with $\mathrm{MaxFE}>$10 lie mostly in this range.
Only 1-4\% of  the {pruned} total have $\mathrm{MaxFE}>$10. 
Over day 1$-$10 both the validity of \texttt{BOXFIT} and the accuracy of \texttt{DeepGlow} improve considerably (Section~\ref{subsec:trdata}).
During real-life detections, estimates should thus be updated daily to benefit from the increased BOXFIT coverage during these first 10 days.

Application to afterglows may be especially relevant 
if the \texttt{BOXFIT}/\texttt{DeepGlow} top-hat jet is preferred for producing a reliable light curve shape and flux.
One reason for this preference could be that the geometry of the transient indicates the jet travels in the direction of the observer, 
as is generally the case in GRBs. 
Even for potentially  more off-axis systems such as gravitational-wave events, \texttt{DeepGlow} can be applicable and even preferred.
In the intermediate phase of the afterglow, physical models (and hence, \texttt{DeepGlow}) outperform semi-analytic ones 
in the prediction of radio light-curve fluxes \citep{Ryan_2020}.
Furthermore, the top-hat jet core dominates the afterglow in these geometries too, once the light curve peaks \citep{2019ApJ...883...15G,duque_radio_2019}.

In cases where a structured jet is required, 
the current version of \texttt{DeepGlow} is not the best choice.

\subsection{Future outlook}
The methods presented here could be extended to emulate  more complex  afterglow models. For GRBs, one example is \texttt{GAMMA}~\citep{ayache_gamma_2021}. It incorporates a precise but highly computationally expensive local cooling approach to the evolution of micro-physical states. Current computational resources reasonably available would not suffice to run such models the roughly $10^5$ times required for accurate emulation using the methods in this work. Instead, a transfer learning approach using \texttt{DeepGlow} as the starting point could prove very valuable and greatly bring down the amount of repeated evaluations required.  

While GRBs are generally observed at boresight, 
the multi-frequency afterglows of gravitational-wave events are generally observed at an angle from the center of the jet.
In these cases, emulating a structured jet would be appropriate. 
These could be based on the numerical hydrodynamical simulations of 
relativistic jets with 
Gaussian profiles \citep{2003ApJ...591.1075K,2022arXiv220707925U}.

\subsection{Open source}
In the current work, we have implemented an 
neural-net emulator for \texttt{BOXFIT}
in \texttt{DeepGlow}, and demonstrated its accuracy. 
To facilitate the inclusion of other, more precise or more applicable RHD models, we have made \texttt{DeepGlow} open source\footnote{\url{https://github.com/OMBoersma/DeepGlow}}, and we encourage contributions. 
We will provide the necessary compute time to train such new or improved emulators. 

\section{CONCLUSION}
\label{sec:conc}
In this work we introduce \texttt{DeepGlow}, an open-source deep learning emulator for the GRB afterglow simulations of \texttt{BOXFIT}. It can generate light curves and spectra to within a few percent accuracy in just a fraction of the \texttt{BOXFIT} evaluation time. It enables rapid characterization of GRB afterglow data using the complex radiative transfer simulations in \texttt{BOXFIT} without the need for a supercomputer. It has support for either an ISM or stellar wind progenitor environment and can be extended to other environments as well.

We estimate the parameters of the broadband GRB afterglow dataset of GRB970508 as a first test of \texttt{DeepGlow}. We find consistent results with an analytical model calibrated to \texttt{BOXFIT} and, in accordance with recent results from the literature, find a decisive preference for a stellar wind progenitor environment around this GRB source.

\begin{acknowledgement}
We thank Mehmet Deniz Aksulu for the extensive discussions on gamma-ray burst physics and Gaussian processes. We thank Hendrik van Eerten for enlightening conversations on \texttt{BOXFIT} and Eliot Ayache for his suggestions which proved instrumental to our deep learning approach. We further thank Daniela Huppenkothen and Anna Watts for comments and suggestions.

This research was supported by the Dutch Research Council (NWO) 
through Vici research program 'ARGO' with project number 639.043.815 and through CORTEX (NWA.1160.18.316), under the research programme NWA-ORC.
\end{acknowledgement}

%% file: appendices.tex
\section{Additional Neural Network Accuracy Figures}
\label{app:extraNNacc}
\begin{figure}[t]
    \includegraphics[width=0.9\textwidth]{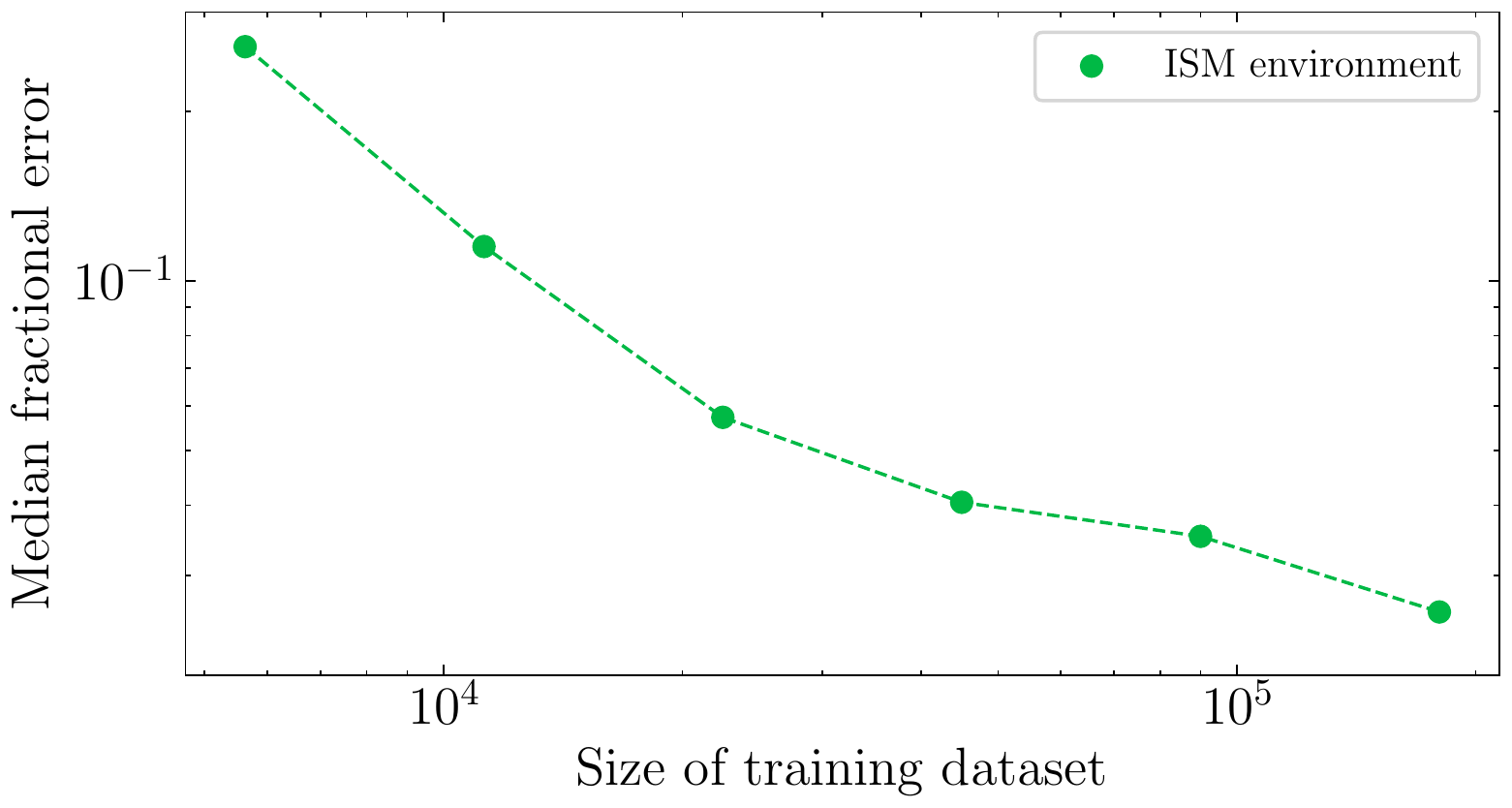}
    \caption{The median fractional error over the test dataset as a function of the training dataset size. The NN was trained for 200 epochs each time.}
    \label{fig:losspertrainingsize}
\end{figure}

In Figure~\ref{fig:losspertrainingsize} we show the median fractional error for the ISM environment NN aggregated over all data points in all light curves in our test dataset as a function of training dataset size. Each NN realisation is trained for 200 epochs. The error follows an approximate log-linear slope. Thus, while the error improves quickly at first, it becomes increasingly harder to increase the accuracy of our NNs by adding more training data.

\begin{figure}[t]
    \includegraphics[width=0.9\textwidth]{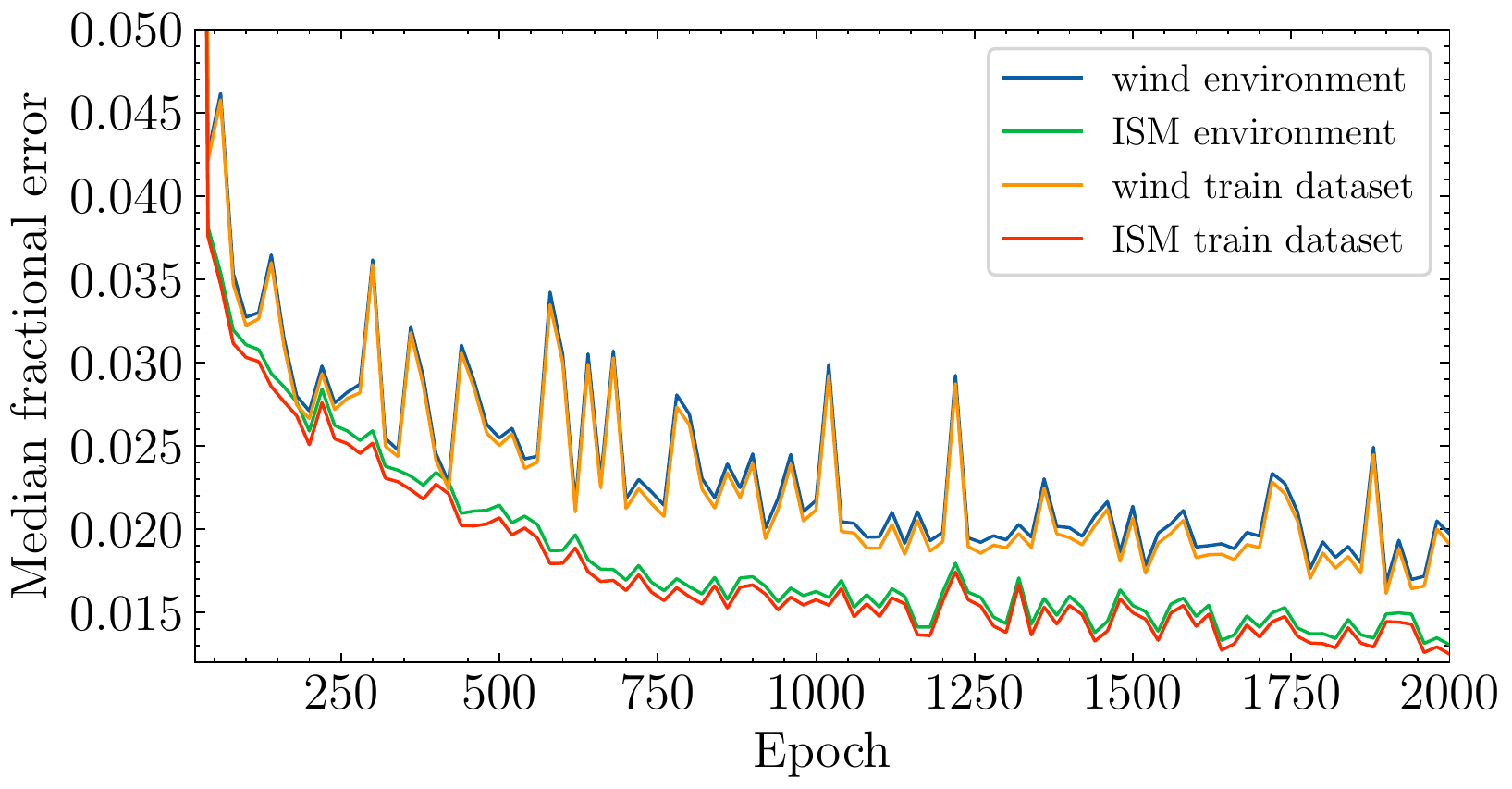}
    \caption{The median fractional error over the train and test dataset as a function of the amount of epochs trained. The ISM environment NN is shown in green (test dataset) or red (train dataset) while the wind environment NN is shown in blue (test dataset) or orange (train dataset).}
    \label{fig:lossperepoch}
\end{figure}

In Figure~\ref{fig:lossperepoch} we show the median fractional error over all data points as a function of the amount of epochs trained. The error decreases rapidly for the first 200 epochs after which it starts to level off to an asymptotic value. The wind environment NN has a much noisier error trajectory compared to the ISM environment NN. This is likely because of the increased amount of light curves with incomplete coverage in the wind environment dataset. Furthermore, the light curves in the wind dataset with incomplete coverage usually have more zero flux values as well.

%% file: pasa-main.bbl
\begin{thebibliography}{}
\providecommand\natexlab[1]{#1}
\providecommand\JournalTitle[1]{#1}

\bibitem[{Abbott {et~al.}(2017{\natexlab{a}})Abbott, Abbott, Abbott, Acernese,
  Ackley, Adams, Adams, Addesso, Adhikari, Adya,
  {et~al.}}]{abbott_gravitational_2017}
Abbott, B.~P., Abbott, R., Abbott, T.~D., {et~al.} 2017{\natexlab{a}},
  \href{http://dx.doi.org/10.3847/2041-8213/aa920c}{\JournalTitle{The
  Astrophysical Journal}, 848, L13}

\bibitem[{Abbott {et~al.}(2017{\natexlab{b}})Abbott, Abbott, Abbott, Acernese,
  Ackley, Adams, Adams, Addesso,
  {et~al.}}]{ligo_scientific_collaboration_and_virgo_collaboration_gw170817_2017}
Abbott, B.~P., Abbott, R., Abbott, T.~D., {et~al.} 2017{\natexlab{b}},
  \href{http://dx.doi.org/10.1103/PhysRevLett.119.161101}{\JournalTitle{Physical
  Review Letters}, 119, 161101}

\bibitem[{Aksulu {et~al.}(2020)Aksulu, Wijers, van Eerten, \& van~der
  Horst}]{aksulu_new_2020}
Aksulu, M.~D., Wijers, R. A. M.~J., van Eerten, H.~J., \& van~der Horst, A.~J.
  2020, \href{http://dx.doi.org/10.1093/mnras/staa2297}{\JournalTitle{Monthly
  Notices of the Royal Astronomical Society}, 497, 4672}

\bibitem[{Aksulu {et~al.}(2022)Aksulu, Wijers, van Eerten, \& van~der
  Horst}]{aksulu_exploring_2022}
Aksulu, M.~D., Wijers, R. A. M.~J., van Eerten, H.~J., \& van~der Horst, A.~J.
  2022, \href{http://dx.doi.org/10.1093/mnras/stac246}{\JournalTitle{Monthly
  Notices of the Royal Astronomical Society}, 511, 2848}

\bibitem[{Alexander {et~al.}(2017)Alexander, Berger, Fong, Williams, Guidorzi,
  Margutti, Metzger, Annis, Blanchard, Brout,
  {et~al.}}]{alexander_electromagnetic_2017}
Alexander, K.~D., Berger, E., Fong, W., {et~al.} 2017,
  \href{http://dx.doi.org/10.3847/2041-8213/aa905d}{\JournalTitle{The
  Astrophysical Journal}, 848, L21}

\bibitem[{Ayache {et~al.}(2021)Ayache, van Eerten, \&
  Eardley}]{ayache_gamma_2021}
Ayache, E.~H., van Eerten, H.~J., \& Eardley, R.~W. 2021,
  \href{http://dx.doi.org/10.1093/mnras/stab3509}{\JournalTitle{Monthly Notices
  of the Royal Astronomical Society}, 510, 1315}

\bibitem[{Blandford \& McKee(1976)}]{blandford_fluid_1976}
Blandford, R.~D. \& McKee, C.~F. 1976,
  \href{http://dx.doi.org/10.1063/1.861619}{\JournalTitle{The Physics of
  Fluids}, 19, 1130}

\bibitem[{{Boersma} \& {van Leeuwen}(2022)}]{2022A&A...664A.160B}
{Boersma}, O.~M. \& {van Leeuwen}, J. 2022,
  \href{http://dx.doi.org/10.1051/0004-6361/202243267}{\JournalTitle{Astronomy
  \& Astrophysics}, 664, A160}

\bibitem[{{Boersma} {et~al.}(2021){Boersma}, {van Leeuwen}, {Adams}, {Adebahr},
  {Kutkin}, {Oosterloo}, {de Blok}, {van den Brink}, {Coolen}, {Connor},
  {et~al.}}]{2021A&A...650A.131B}
{Boersma}, O.~M., {van Leeuwen}, J., {Adams}, E.~A.~K., {et~al.} 2021,
  \href{http://dx.doi.org/10.1051/0004-6361/202140578}{\JournalTitle{Astronomy
  \& Astrophysics}, 650, A131}

\bibitem[{Buchner {et~al.}(2014)Buchner, Georgakakis, Nandra, Hsu, Rangel,
  Brightman, Merloni, Salvato, Donley, \& Kocevski}]{buchner_x-ray_2014}
Buchner, J., Georgakakis, A., Nandra, K., {et~al.} 2014,
  \href{http://dx.doi.org/10.1051/0004-6361/201322971}{\JournalTitle{Astronomy
  \& Astrophysics}, 564, A125}

\bibitem[{Caruana {et~al.}(2000)Caruana, Lawrence, \&
  Giles}]{caruana_overfitting_2000}
Caruana, R., Lawrence, S., \& Giles, L. 2000, in Proceedings of the 13th
  {International} {Conference} on {Neural} {Information} {Processing}
  {Systems}, {NIPS}'00 (Cambridge, MA, USA: MIT Press), 381

\bibitem[{Chornock {et~al.}(2017)Chornock, Berger, Kasen, Cowperthwaite,
  Nicholl, Villar, Alexander, Blanchard, Eftekhari, Fong,
  {et~al.}}]{chornock_electromagnetic_2017}
Chornock, R., Berger, E., Kasen, D., {et~al.} 2017,
  \href{http://dx.doi.org/10.3847/2041-8213/aa905c}{\JournalTitle{The
  Astrophysical Journal}, 848, L19}

\bibitem[{Costa {et~al.}(1997)Costa, Frontera, Heise, Feroci, in~'t Zand,
  Fiore, Cinti, Dal~Fiume, Nicastro, Orlandini,
  {et~al.}}]{costa_discovery_1997}
Costa, E., Frontera, F., Heise, J., {et~al.} 1997,
  \href{http://dx.doi.org/10.1038/42885}{\JournalTitle{Nature}, 387, 783}

\bibitem[{Coulter {et~al.}(2017)Coulter, Foley, Kilpatrick, Drout, Piro,
  Shappee, Siebert, Simon, Ulloa, Kasen, {et~al.}}]{coulter_swope_2017}
Coulter, D.~A., Foley, R.~J., Kilpatrick, C.~D., {et~al.} 2017,
  \href{http://dx.doi.org/10.1126/science.aap9811}{\JournalTitle{Science}, 358,
  1556}

\bibitem[{Cybenko(1989)}]{cybenko_approximation_1989}
Cybenko, G. 1989,
  \href{http://dx.doi.org/10.1007/BF02551274}{\JournalTitle{Mathematics of
  Control, Signals and Systems}, 2, 303}

\bibitem[{Duque {et~al.}(2019)Duque, Daigne, \& Mochkovitch}]{duque_radio_2019}
Duque, R., Daigne, F., \& Mochkovitch, R. 2019,
  \href{http://dx.doi.org/10.1051/0004-6361/201935926}{\JournalTitle{Astronomy
  \& Astrophysics}, 631, A39}

\bibitem[{Eichler \& Waxman(2005)}]{eichler_efficiency_2005}
Eichler, D. \& Waxman, E. 2005,
  \href{http://dx.doi.org/10.1086/430596}{\JournalTitle{The Astrophysical
  Journal}, 627, 861}

\bibitem[{Feroz {et~al.}(2009)Feroz, Hobson, \& Bridges}]{feroz_multinest_2009}
Feroz, F., Hobson, M.~P., \& Bridges, M. 2009,
  \href{http://dx.doi.org/10.1111/j.1365-2966.2009.14548.x}{\JournalTitle{Monthly
  Notices of the Royal Astronomical Society}, 398, 1601}

\bibitem[{Gao {et~al.}(2013)Gao, Lei, Zou, Wu, \& Zhang}]{gao_complete_2013}
Gao, H., Lei, W.-H., Zou, Y.-C., Wu, X.-F., \& Zhang, B. 2013,
  \href{http://dx.doi.org/10.1016/j.newar.2013.10.001}{\JournalTitle{New
  Astronomy Reviews}, 57, 141}

\bibitem[{{Gehrels} {et~al.}(2004){Gehrels}, {Chincarini}, {Giommi}, {Mason},
  {Nousek}, {Wells}, {White}, {Barthelmy}, {Burrows}, {Cominsky}, {Hurley},
  {Marshall}, {M{\'e}sz{\'a}ros}, {Roming}, {Angelini}, {Barbier}, {Belloni},
  {Campana}, {Caraveo}, {Chester}, {Citterio}, {Cline}, {Cropper}, {Cummings},
  {Dean}, {Feigelson}, {Fenimore}, {Frail}, {Fruchter}, {Garmire}, {Gendreau},
  {Ghisellini}, {Greiner}, {Hill}, {Hunsberger}, {Krimm}, {Kulkarni}, {Kumar},
  {Lebrun}, {Lloyd-Ronning}, {Markwardt}, {Mattson}, {Mushotzky}, {Norris},
  {Osborne}, {Paczynski}, {Palmer}, {Park}, {Parsons}, {Paul}, {Rees},
  {Reynolds}, {Rhoads}, {Sasseen}, {Schaefer}, {Short}, {Smale}, {Smith},
  {Stella}, {Tagliaferri}, {Takahashi}, {Tashiro}, {Townsley}, {Tueller},
  {Turner}, {Vietri}, {Voges}, {Ward}, {Willingale}, {Zerbi}, \&
  {Zhang}}]{gehrels_swift_2004}
{Gehrels}, N., {Chincarini}, G., {Giommi}, P., {et~al.} 2004,
  \href{http://dx.doi.org/10.1086/422091}{\JournalTitle{The Astrophysical
  Journal}, 611, 1005}

\bibitem[{{Gill} {et~al.}(2019){Gill}, {Granot}, {De Colle}, \&
  {Urrutia}}]{2019ApJ...883...15G}
{Gill}, R., {Granot}, J., {De Colle}, F., \& {Urrutia}, G. 2019,
  \href{http://dx.doi.org/10.3847/1538-4357/ab3577}{\JournalTitle{The
  Astrophysical Journal}, 883, 15}

\bibitem[{Granot \& Sari(2002)}]{granot_shape_2002}
Granot, J. \& Sari, R. 2002,
  \href{http://dx.doi.org/10.1086/338966}{\JournalTitle{The Astrophysical
  Journal}, 568, 820}

\bibitem[{Haggard {et~al.}(2017)Haggard, Nynka, Ruan, Kalogera, Cenko, Evans,
  \& Kennea}]{haggard_deep_2017}
Haggard, D., Nynka, M., Ruan, J.~J., {et~al.} 2017,
  \href{http://dx.doi.org/10.3847/2041-8213/aa8ede}{\JournalTitle{The
  Astrophysical Journal}, 848, L25}

\bibitem[{Hallinan {et~al.}(2017)Hallinan, Corsi, Mooley, Hotokezaka, Nakar,
  Kasliwal, Kaplan, Frail, Myers, Murphy, {et~al.}}]{hallinan_radio_2017}
Hallinan, G., Corsi, A., Mooley, K.~P., {et~al.} 2017,
  \href{http://dx.doi.org/10.1126/science.aap9855}{\JournalTitle{Science}, 358,
  1579}

\bibitem[{Higgins {et~al.}(2019)Higgins, van~der Horst, Starling, Anderson,
  Perley, van Eerten, Wiersema, Jakobsson, Kouveliotou, Lamb,
  {et~al.}}]{higgins_detailed_2019}
Higgins, A.~B., van~der Horst, A.~J., Starling, R. L.~C., {et~al.} 2019,
  \href{http://dx.doi.org/10.1093/mnras/stz384}{\JournalTitle{Monthly Notices
  of the Royal Astronomical Society}, 484, 5245}

\bibitem[{Hornik {et~al.}(1989)Hornik, Stinchcombe, \&
  White}]{hornik_multilayer_1989}
Hornik, K., Stinchcombe, M., \& White, H. 1989,
  \href{http://dx.doi.org/10.1016/0893-6080(89)90020-8}{\JournalTitle{Neural
  Networks}, 2, 359}

\bibitem[{Kasim {et~al.}(2021)Kasim, Watson-Parris, Deaconu, Oliver, Hatfield,
  Froula, Gregori, Jarvis, Khatiwala, Korenaga, {et~al.}}]{kasim_building_2021}
Kasim, M.~F., Watson-Parris, D., Deaconu, L., {et~al.} 2021,
  \href{http://dx.doi.org/10.1088/2632-2153/ac3ffa}{\JournalTitle{Machine
  Learning: Science and Technology}, 3, 015013}

\bibitem[{Kass \& Raftery(1995)}]{kass_bayes_1995}
Kass, R.~E. \& Raftery, A.~E. 1995,
  \href{http://dx.doi.org/10.1080/01621459.1995.10476572}{\JournalTitle{Journal
  of the American Statistical Association}, 90, 773}

\bibitem[{Kerzendorf {et~al.}(2021)Kerzendorf, Vogl, Buchner, Contardo,
  Williamson, \& Smagt}]{kerzendorf_dalek_2021}
Kerzendorf, W.~E., Vogl, C., Buchner, J., {et~al.} 2021,
  \href{http://dx.doi.org/10.3847/2041-8213/abeb1b}{\JournalTitle{The
  Astrophysical Journal Letters}, 910, L23}

\bibitem[{{Kumar} \& {Granot}(2003)}]{2003ApJ...591.1075K}
{Kumar}, P. \& {Granot}, J. 2003,
  \href{http://dx.doi.org/10.1086/375186}{\JournalTitle{The Astrophysical
  Journal}, 591, 1075}

\bibitem[{{Leventis} {et~al.}(2012){Leventis}, {van Eerten}, {Meliani}, \&
  {Wijers}}]{2012MNRAS.427.1329L}
{Leventis}, K., {van Eerten}, H.~J., {Meliani}, Z., \& {Wijers}, R.~A.~M.~J.
  2012,
  \href{http://dx.doi.org/10.1111/j.1365-2966.2012.21994.x}{\JournalTitle{Monthly
  Notices of the Royal Astronomical Society}, 427, 1329}

\bibitem[{Nousek {et~al.}(2006)Nousek, Kouveliotou, Grupe, Page, Granot,
  Ramirez-Ruiz, Patel, Burrows, Mangano, Barthelmy, Beardmore, Campana,
  Capalbi, Chincarini, Cusumano, Falcone, Gehrels, Giommi, Goad, Godet,
  Hurkett, Kennea, Moretti, O’Brien, Osborne, Romano, Tagliaferri, \&
  Wells}]{nousek_evidence_2006}
Nousek, J.~A., Kouveliotou, C., Grupe, D., {et~al.} 2006,
  \href{http://dx.doi.org/10.1086/500724}{\JournalTitle{The Astrophysical
  Journal}, 642, 389}

\bibitem[{Panaitescu \& Kumar(2002)}]{panaitescu_properties_2002}
Panaitescu, A. \& Kumar, P. 2002,
  \href{http://dx.doi.org/10.1086/340094}{\JournalTitle{The Astrophysical
  Journal}, 571, 779}

\bibitem[{Pedregosa {et~al.}(2011)Pedregosa, Varoquaux, Gramfort, Michel,
  Thirion, Grisel, Blondel, Prettenhofer, Weiss, Dubourg, Vanderplas, Passos,
  Cournapeau, Brucher, Perrot, \& Duchesnay}]{scikit-learn}
Pedregosa, F., Varoquaux, G., Gramfort, A., {et~al.} 2011,
  \JournalTitle{Journal of Machine Learning Research}, 12, 2825

\bibitem[{Rasmussen \& Williams(2006)}]{rasmussen_gaussian_2006}
Rasmussen, C.~E. \& Williams, C. K.~I. 2006, Gaussian processes for machine
  learning, Adaptive computation and machine learning (Cambridge, Mass: MIT
  Press)

\bibitem[{Ryan {et~al.}(2015)Ryan, Eerten, MacFadyen, \&
  Zhang}]{ryan_gamma-ray_2015}
Ryan, G., Eerten, H.~v., MacFadyen, A., \& Zhang, B.-B. 2015,
  \href{http://dx.doi.org/10.1088/0004-637x/799/1/3}{\JournalTitle{The
  Astrophysical Journal}, 799, 3}

\bibitem[{Ryan {et~al.}(2020)Ryan, van Eerten, Piro, \& Troja}]{Ryan_2020}
Ryan, G., van Eerten, H., Piro, L., \& Troja, E. 2020,
  \href{http://dx.doi.org/10.3847/1538-4357/ab93cf}{\JournalTitle{The
  Astrophysical Journal}, 896, 166}

\bibitem[{Sari {et~al.}(1998)Sari, Piran, \& Narayan}]{sari_spectra_1998}
Sari, R., Piran, T., \& Narayan, R. 1998,
  \href{http://dx.doi.org/10.1086/311269}{\JournalTitle{The Astrophysical
  Journal}, 497, L17}

\bibitem[{Schmidhuber(2015)}]{schmidhuber_deep_2015}
Schmidhuber, J. 2015,
  \href{http://dx.doi.org/10.1016/j.neunet.2014.09.003}{\JournalTitle{Neural
  Networks}, 61, 85}

\bibitem[{Schmit \& Pritchard(2018)}]{schmit_emulation_2018}
Schmit, C.~J. \& Pritchard, J.~R. 2018,
  \href{http://dx.doi.org/10.1093/mnras/stx3292}{\JournalTitle{Monthly Notices
  of the Royal Astronomical Society}, 475, 1213}

\bibitem[{Smith(2015)}]{smith_cyclical_2015}
Smith, L.~N. 2015,
  \href{http://dx.doi.org/10.48550/arXiv.1506.01186}{\JournalTitle{arXiv
  e-prints}, arXiv:1506.01186}

\bibitem[{Sutskever {et~al.}(2013)Sutskever, Martens, Dahl, \&
  Hinton}]{sutskever_importance_2013}
Sutskever, I., Martens, J., Dahl, G., \& Hinton, G. 2013,
  \href{https://proceedings.mlr.press/v28/sutskever13.html}{in Proceedings of
  the 30th {International} {Conference} on {Machine} {Learning}} (Pmlr), 1139

\bibitem[{{Urrutia} {et~al.}(2022){Urrutia}, {De Colle}, \&
  {L{\'o}pez-C{\'a}mara}}]{2022arXiv220707925U}
{Urrutia}, G., {De Colle}, F., \& {L{\'o}pez-C{\'a}mara}, D. 2022,
  \JournalTitle{arXiv e-prints}, arXiv:2207.07925

\bibitem[{Van~Eerten {et~al.}(2010)Van~Eerten, Leventis, Meliani, Wijers, \&
  Keppens}]{van_eerten_gamma-ray_2010}
Van~Eerten, H.~J., Leventis, K., Meliani, Z., Wijers, R., \& Keppens, R. 2010,
  \JournalTitle{Monthly Notices of the Royal Astronomical Society}, 403, 300

\bibitem[{van Eerten {et~al.}(2012)van Eerten, van~der Horst, \&
  MacFadyen}]{van_eerten_gamma-ray_2012}
van Eerten, H.~J., van~der Horst, A., \& MacFadyen, A. 2012,
  \href{http://dx.doi.org/10.1088/0004-637x/749/1/44}{\JournalTitle{The
  Astrophysical Journal}, 749, 44}

\bibitem[{{van Leeuwen} {et~al.}(2023){van Leeuwen}, {Kooistra}, {Oostrum},
  {Connor}, {Hargreaves}, {Maan}, {Pastor-Marazuela}, {Petroff}, {van der
  Schuur}, {Sclocco}, {Straal}, {Vohl}, {Wijnholds}, {Adams}, {Adebahr},
  {Attema}, {Bassa}, {Bast}, {Bilous}, {de Blok}, {Boersma}, {van Cappellen},
  {Coolen}, {Damstra}, {D{\'e}nes}, {van Diepen}, {Gardenier}, {Grange},
  {Gunst}, {Hess}, {Holties}, {van der Hulst}, {Hut}, {Kutkin}, {Loose},
  {Lucero}, {Mika}, {Mikhailov}, {Morganti}, {Moss}, {Mulder}, {Norden},
  {Oosterloo}, {Orr{\'u}}, {Paragi}, {de Reijer}, {Schoenmakers}, {Stuurwold},
  {ter Veen}, {Wang}, {Zanting}, \& {Ziemke}}]{van_leeuwen_apertif_2022}
{van Leeuwen}, J., {Kooistra}, E., {Oostrum}, L., {et~al.} 2023,
  \href{http://dx.doi.org/10.1051/0004-6361/202244107}{\JournalTitle{Astronomy
  \& Astrophysics}, 672, A117}

\bibitem[{Wijers {et~al.}(1997)Wijers, Rees, \&
  M\'esz\'aros}]{wijers_shocked_1997}
Wijers, R. A. M.~J., Rees, M.~J., \& M\'esz\'aros, P. 1997,
  \href{http://dx.doi.org/10.1093/mnras/288.4.L51}{\JournalTitle{Monthly
  Notices of the Royal Astronomical Society}, 288, L51}

\bibitem[{Ying(2019)}]{ying_overview_2019}
Ying, X. 2019,
  \href{http://dx.doi.org/10.1088/1742-6596/1168/2/022022}{\JournalTitle{Journal
  of Physics: Conference Series}, 1168, 022022}

\bibitem[{Yost {et~al.}(2003)Yost, Harrison, Sari, \& Frail}]{yost_study_2003}
Yost, S.~A., Harrison, F.~A., Sari, R., \& Frail, D.~A. 2003,
  \href{http://dx.doi.org/10.1086/378288}{\JournalTitle{The Astrophysical
  Journal}, 597, 459}

\end{thebibliography}
